\documentclass[]{emulateapj}
\shorttitle{YSO carbon isotope partitioning}
\shortauthors{Smith et al.}
\usepackage{multirow}

\begin{document}

\title{Heterogeneity in $^{12}$CO/$^{13}$CO Ratios Toward Solar-type Young Stellar Objects}

\author{Rachel L. Smith\altaffilmark{1,2}, Klaus M. Pontoppidan\altaffilmark{3}, Edward D. Young\altaffilmark{4} and
Mark R. Morris\altaffilmark{5}}
\altaffiltext{1} {North Carolina Museum of Natural Sciences, 121 West Jones Street, Raleigh, NC, 27603; rachel.smith@naturalsciences.org}
\altaffiltext{2} {Department of Physics and Astronomy, Appalachian State University, 525 Rivers Street, Boone, NC 28608-2106; smithrl2@appstate.edu}
\altaffiltext{3} {Space Telescope Science Institute, 3700 San Martin Drive, Baltimore, MD 21218}
\altaffiltext{4} {Department of Earth and Space Sciences, University of California Los Angeles, 595 Charles E. Young Drive East, Geology Building, Los Angeles, CA 90095-1567}
\altaffiltext{5} {Division of Astronomy and Astrophysics, Department of Physics and Astronomy, University of California, Los Angeles, CA 90095-1547}

\begin{abstract}
This study reports an unusual heterogeneity in [$^{12}$C$^{16}$O]/[$^{13}$C$^{16}$O] abundance ratios of carbon monoxide observed in the gas phase toward seven $\sim$ solar-mass YSOs and three dense foreground clouds in the nearby star-forming regions, Ophiuchus, Corona Australis, Orion, Vela and an isolated core, L43. Robust isotope ratios were derived using infrared absorption spectroscopy of the 4.7\,$\mu$m fundamental and 2.3\,$\mu$m overtone rovibrational bands of CO at very high resolution ($\lambda$/$\Delta$$\lambda\approx 95,000$), observed with the CRIRES spectrograph on the Very Large Telescope. We find [$^{12}$C$^{16}$O]/[$^{13}$C$^{16}$O] values ranging from $\sim$ 85 to 165, significantly higher than those of the local interstellar medium ($\sim 65$ to 69). These observations are evidence for isotopic heterogeneity in carbon reservoirs in solar-type YSO environments, and encourage the need for refined Galactic chemical evolution models to explain the $^{12}$C/$^{13}$C discrepancy between the solar system and local ISM. The oxygen isotope ratios are consistent with isotopologue-specific photodissociation by CO self-shielding toward the disks, VV CrA N and HL Tau, further substantiating models predicting CO self-shielding on disk surfaces. However, we find that CO self-shielding is an unlikely general explanation for the high [$^{12}$C$^{16}$O]/[$^{13}$C$^{16}$O] ratios observed in this study. Comparison of the solid CO against gas-phase [$^{12}$C$^{16}$O]/[$^{13}$C$^{16}$O] suggests that interactions between CO ice and gas reservoirs need to be further investigated as at least a partial explanation for the unusually high [$^{12}$C$^{16}$O]/[$^{13}$C$^{16}$O] observed.
\end{abstract}


\section{Introduction}
Observations of carbon monoxide (CO) provide a robust window into circumstellar chemistry. Due to the large percentage of carbon ($\sim15$ to $30\%$) and oxygen ($\sim10\%$) tied up in CO in the early stages of young stellar object (YSO) formation \citep{dickman1978}, CO is a valuable tracer of chemical inheritance from the envelope to the disk in solar-type YSOs. The use of high-resolution spectroscopy to observe CO at infrared wavelengths enables high-precision investigations of isotope fractionation in both oxygen and carbon toward YSOs in embedded and disk-phase evolutionary stages, and is consequently an important tool for connecting the solar system to currently forming protoplanetary systems. Examples include observational studies toward diffuse clouds and inferring isotopologue-specific photodissociation by CO self-shielding as a potentially important mechanism of carbon isotope fractionation in CO in the interstellar medium \citep{Bally1982, sheffer2002}. Further, high-resolution near-infrared observations of the CO rovibrational band toward protoplanetary disks \citep{brittain2005, smith2009} and protostellar envelopes have recently revealed signatures in CO isotopologue ratios that are consistent with CO self-shielding, with precision sufficiently high for comparison to meteoritic data \citep{smith2009}.

Observations of isotope ratios toward YSOs and, where possible, their parent clouds, are valuable in assessing theorized similarities between circumstellar envelopes, disks and dense clouds local to their origin. This is important not only for evaluating predictions of galactic chemical evolution (GCE) models that are constrained by solar system abundances, but also for characterizing the range of carbon isotope ratios toward YSOs varying with evolutionary stage, and establishing accurate astrochemical constraints on protoplanetary environments. Observational studies can help demonstrate whether there are significant isotopic differences or trends between diffuse clouds, dense molecular clouds, and YSOs. These in turn help construct explanations for apparent chemical anomalies, such as the $^{12}$C/$^{13}$C solar system-interstellar medium (ISM) discrepancy and solar system oxygen isotope anomaly. 

This paper focuses on the carbon isotopic systems in YSOs as traced by high-resolution observations of CO isotopologues, with additional mention of significant findings in oxygen isotopes. Current measurements of the $^{12}$C/$^{13}$C ratio for the solar system are $\sim86$ to 89, derived from photospheric models applied to the solar CO spectrum \citep{scott2006}, and bulk solar system materials \citep{Rosman1998, Wilson1999, coplen2002, clayton2004, young2011}. Values of $^{12}$C/$^{13}$C for the solar system have been consistently higher than those of $\sim62$ to 69 observed toward the local ISM, measured using millimeter to sub-millimeter rotational transitions of various carbon-bearing molecules, CO, CN, CH$^{+}$,  and H$_{2}$CO \citep{langer1993, Wilson1999, Milam2005, casassus2005}. The apparent discrepancy in $^{12}$C/$^{13}$C between the solar system and ISM has been attributed to GCE leading to enrichment of $^{13}$C relative to $^{12}$C over the $\sim4.6$ Gyr lifetime of the solar system \citep[e.g.][]{Milam2005, Prantzos1996}. An important property of GCE models is the assumption that the solar system records the chemical state of the molecular cloud material $\sim4.6$\,Gyr ago. That is, the assumption that the solar nebula was isotopically representative of its parent cloud at its Galactic radius at the time of formation should also hold for present-day solar-type YSOs.

The challenge for observational comparisons of solar system isotope ratios in general is the need for high precision in molecular abundance measurements. While techniques using strong rotational lines in the millimeter/sub-millimeter range often involve model-dependent comparisons between optically thick and thin lines, high-resolution infrared spectrographs have enabled significant breakthroughs in observations of nearby YSOs. Infrared absorption spectroscopy of optically thin CO lines allows for precise column density measurements along pencil-beam lines-of-sight, enabling the derivation of precise column densities and isotope ratios from the line profiles of as many as four currently observable CO isotopologues ($^{12}$C$^{16}$O, $^{13}$C$^{16}$O, $^{12}$C$^{18}$O, $^{12}$C$^{17}$O). Robust isotope ratios so derived can be compared with statistical significance to sub-millimeter ISM values, solar abundances, and meteoritic oxygen. A high degree of precision (within $10\%$) on the absolute values of the column densities has been demonstrated for the four most common CO isotopologues \citep{smith2009}. The study reported here builds on this work. 

Our targets span a range in $^{12}$C$^{16}$O ice optical depth, from nearly saturated to undetected, providing an opportunity to investigate a potential observational connection between the CO ice and gas reservoirs by comparing the gas-phase [$^{12}$C$^{16}$O]/[$^{13}$C$^{16}$O] to the CO ice column along a single line-of-sight. Since solids tend to concentrate the heavy isotope, relative ice-gas partioning may lead to a paucity of $^{13}$C (or excess of $^{12}$C) in the gas phase, with the opposite signature sequestered in the ice.

\section{Observations} 
Data for all targets were obtained as part of a European Southern Observatory Large Program \citep{Pontoppidan2011b} to observe $\sim 100$ YSOs and protoplanetary disks with the Cryogenic Infrared Echelle Spectrograph (CRIRES) at the Very Large Telescope (VLT) in Chile. The CRIRES instrument is an adaptive optics (AO)-assisted spectrometer that operates at very high resolving power ($\lambda$/$\Delta$$\lambda$ $\approx$ 95,000). 

\begin{deluxetable}{llll}[ht]
\centering
\tablecolumns{4}
\tablewidth{0pt}
\tablecaption{Summary of Observed Targets}

\tablehead{
\colhead{Target} & \colhead{Region\tablenotemark{a}} & \colhead{Distance} & \colhead{Comments\tablenotemark{a}}\\
       &                                      & \colhead{(pc)}     &                  
   }
\startdata
DoAr24E S,N                & Oph & 125 & Stage II/IR companion \\
IRS 43                  & Oph & 125 & Stage I\\
IRS 51                  & Oph & 125 & Extincted stage II\\
IRS 63\tablenotemark{b} & Oph & 125 & Stage I\\
RNO 91                  & L43 & 125 & Stage I\\
VV CrA N,S                  & CrA & 130 & Stage II/IR companion \\
HL Tau                  & Tau & 140 & Stage I\\
Reipurth 50             & Ori & 470 & Stage I/FU Ori \\
LLN 19                   & Vel & 700 & Stage I
\enddata
\tablecomments{
\tablenotetext{a}{The evolutionary stage refers to the system proposed by \cite{robitaille2006}. The two targets with N/S components are 1-2\arcsec binaries, for which the secondary is a highly extincted ``infrared companion'' \citep{koresko1997}.}
\tablenotetext{b}{IRS 63 may be formed from an isolated core \citep{ridge2006}.}
}
\label{Target_table}
\end{deluxetable}

\subsection{Strategy}
Targets tracing high column densities of molecular gas were selected based on evidence of deep CO absorption lines at 4.7\,$\mu$m in spectra obtained with VLT-ISAAC \citep{Pontoppidan2003} and Keck-NIRSPEC \citep[e.g.,][]{boogert2002b}. Selected targets were observed by CRIRES in the CO overtone ($v=2-0$) band at 2.3 $\mu$m and the fundamental ($v=1-0$) band at 4.7 $\mu$m. Because the overtone transitions are weaker than those of the fundamental band by a factor of $\sim100$, the former are typically optically thin for $^{12}$C$^{16}$O, thereby enabling the precise measurement of column densities for this molecule. The spectra of the fundamental rovibrational bands were obtained using the $0\farcs2$ slit, resulting in a resolving power of $R=\lambda/\Delta\lambda \approx 95,000$ (corresponding to 3.2\,km\,s$^{-1}$). Several settings in the fundamental band were obtained for all targets in order to probe a range of rotational levels for $^{13}$C$^{16}$O, $^{12}$C$^{18}$O and $^{12}$C$^{17}$O, spanning $J = 0$ to at least $J=9$. Observations were timed such that the Doppler shifts of the targets relative to the Earth's atmosphere were at least 20\,km\,s$^{-1}$ due to a combination of the intrinsic target velocity and reflex motion of the Earth. As a consequence, telluric CO lines are typically not affecting the absorption lines of the targets.
\begin{deluxetable*}{lllll}[ht]
\centering
\tablecolumns{5}
\tablewidth{0pt}
\tablecaption{Summary of Observed Targets}

\tablehead{
\colhead{Target}  &\colhead{RA} &\colhead{Dec}  &\colhead{Spectral Range}  &\colhead{Obs. Date} \\
  &\colhead{(J2000)} &\colhead{(J2000)}  &\colhead{($\mu$m)}  &\colhead{(UT)}      
     }
\startdata
DoAr24E S,N\tablenotemark{a} & 16 26 23.36 & -24 20 59.8 & $4.644-4.768$ & 2007-09-02 \\
                         &             &             & $2.338-2.396$ & 2010-03-18 \\
IRS 43                   & 16 27 26.94 & -24 40 50.8 & $4.645-4.900$ & 2008-08-05 \\
                         &             &             & $2.338-2.396$ & 2008-08-04 \\
IRS 51                   & 16 27 39.83 & -24 43 15.1 & $4.645-4.768$ & 2008-08-02 \\
                         &             &             & $2.338-2.391$ & 2008-08-03 \\
IRS 63\tablenotemark{a}  & 16 31 35.65 & -24 01 29.3 & $4.645-4.978$ & 2007-04-25 \\
                         &             &             & $2.338-2.396$ & 2008-08-07 \\
RNO 91                   & 16 34 29.28 & -15 47 01.5 & $4.645-4.855$ & 2010-03-18 \\
                        &             &             & $2.338-2.396$ & 2010-04-08 \\
VV CrA N,S                 & 19 03 06.74 & -37 12 49.7 & $4.701-4.815$ & 2007-08-31 \\
                         &             &             & $2.338-2.396$ & 2007-08-31 \\
HL Tau                   & 04 31 38.44 &  18 13 57.7 & $4.643-4.853$ & 2010-01-30 \\
                         &             &             & $2.337-2.396$ & 2010-01-28 \\
Reipurth 50              & 05 40 27.70 & -07 27 28.0 & $4.645-4.901$ & 2007-10-11 \\
                         &             &             & $2.338-2.396$ & 2007-10-17 \\
LLN 19                    & 08 48 48.00 & -43 32 25.0 & $4.644-4.867$ & 2007-04-25 \\
                       &             &             & $2.253-2.355$ & 2007-04-25 
\enddata
\tablecomments{
\tablenotetext{a}{The non-coevality of these observations may contribute some level of uncertainty to the derived isotopic abundances, as discussed in the text.}
}
\label{Obs_table}
\end{deluxetable*}

Table \ref{Target_table} summarizes the region, distance and evolutionary stage of the observed targets, listed in order of increasing distance from the local solar neighborhood ($R_{\rm GC}\sim8$\,kpc). Specific observing parameters are listed in Table \ref{Obs_table}.
Selected targets probe a range of circumstellar environments, as noted. Because the lines-of-sight traced by infrared absorption may pass through the upper layers of an inclined disk as well as through envelope material and a foreground molecular cloud unrelated to the YSO, some portion of each of these environments may be represented in the integrated spectra, depending on the target. For instance, near-infrared absorption spectra observed toward the extincted protoplanetary disk, IRS 51, will be an integrated sight line consisting mostly of the foreground Ophiuchus ridge and a relatively small contribution of the protoplanetary disk. This geometry has been predicted by interferometric observations \citep{vankempen2009} and is supported by our observations that reveal cold-only gas lines.

\begin{figure}[ht]
\centering
\includegraphics[width=8.5cm]{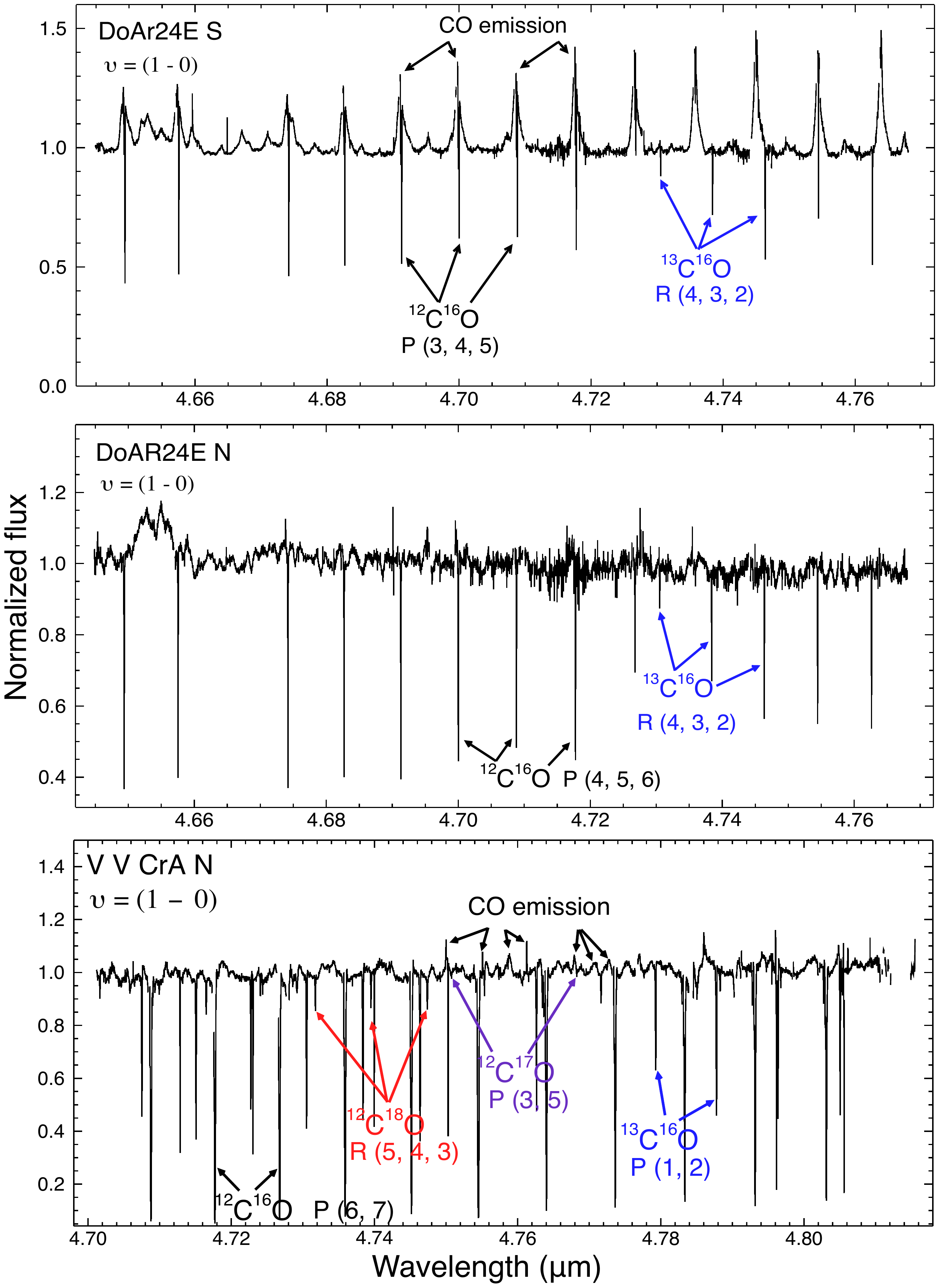}
\caption[]{Infrared absorption spectra of the CO fundamental rovibrational band toward the DoAr24E N,S binary and the disk VV CrA N, which have weak CO ice optical depth upper limits (Table \ref{ice_table}). Spectra for DoAr24E N,S show few CO absorption lines, while VV CrA N shows a rich CO forest of narrow rovibrational absorption lines. Sample lines are marked. The low-level emission from hot CO gas toward DoAr24 S and VV CrA N is possibly representative of inner disk disk material. The weak emission line at $\sim4.65\,\mu$m for DoAR24E N is HI Pfund$\beta$. The main portions of the spectra used for the line analyses are presented.}
\label{specm1a}
\end{figure}

\subsection{Data reduction}
The CRIRES spectra were reduced using standard procedures for infrared echelle spectroscopy, including flat-field correction, adjustments to account for detector non-linearity, and linearization of the spectral traces in dispersion- and cross-dispersion directions. The spectra were optimally extracted, wavelength-calibrated using the telluric absorption lines referenced to an atmospheric model spectrum, and transformed to the local standard-of-rest frame. Relative flux calibration was carried out by dividing the targets by observations of early-type telluric standards observed immediately surrounding observations of the science target at the same airmass. The division was optimized by modifying the telluric spectra to account for any difference in grating angle, airmass, and spectral resolving power. To aid in the measurement of any intrinsic line broadening, the resolving powers for each science spectrum were measured from telluric spectra to within an accuracy of $\sim0.1$\,km\,s$^{-1}$, by fitting the ``Reference Forward Model'' (RFM\footnote{\tt http://www.atm.ox.ac.uk/RFM}, see also Section \ref{sec:derivation}). Further details on CRIRES data processing is described in \cite{Pontoppidan2008, pontoppidan2011a}.

\begin{figure}[ht]
\centering
\includegraphics[width=8.5cm]{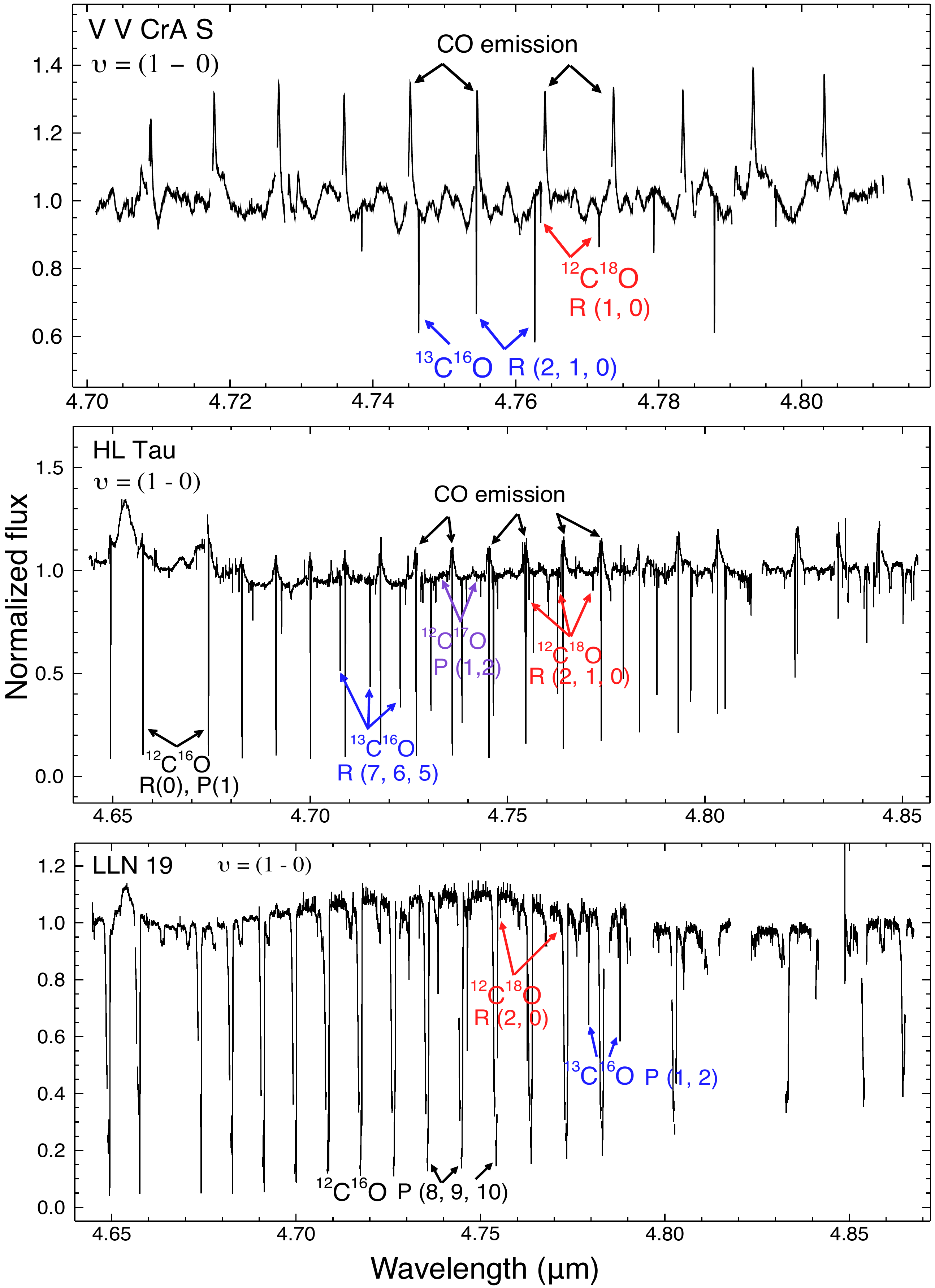}
\caption[]{Infrared absorption spectra of the CO fundamental rovibrational band toward VV CrA S, HL Tau and LLN 19, characterized by weak lower-limits (VV CrA S and HL Tau), or shallow (LLN 19) CO ice depths (Table \ref{ice_table}). HL Tau and LLN 19 have a rich CO forest of narrow rovibrational absorption lines due to the various CO isotopologues present. Sample lines are marked. Few lines of the rarer isotopologues ($^{12}$C$^{18}$O and $^{12}$C$^{17}$O) are seen toward the more deeply embedded object, LLN 19. Low-level emission toward VV CrA S and HL Tau are possibly representative of hot CO gas of the inner disk.}
\label{specm1b}
\end{figure}

\begin{figure}[ht]
\centering
\includegraphics[width=8.5cm]{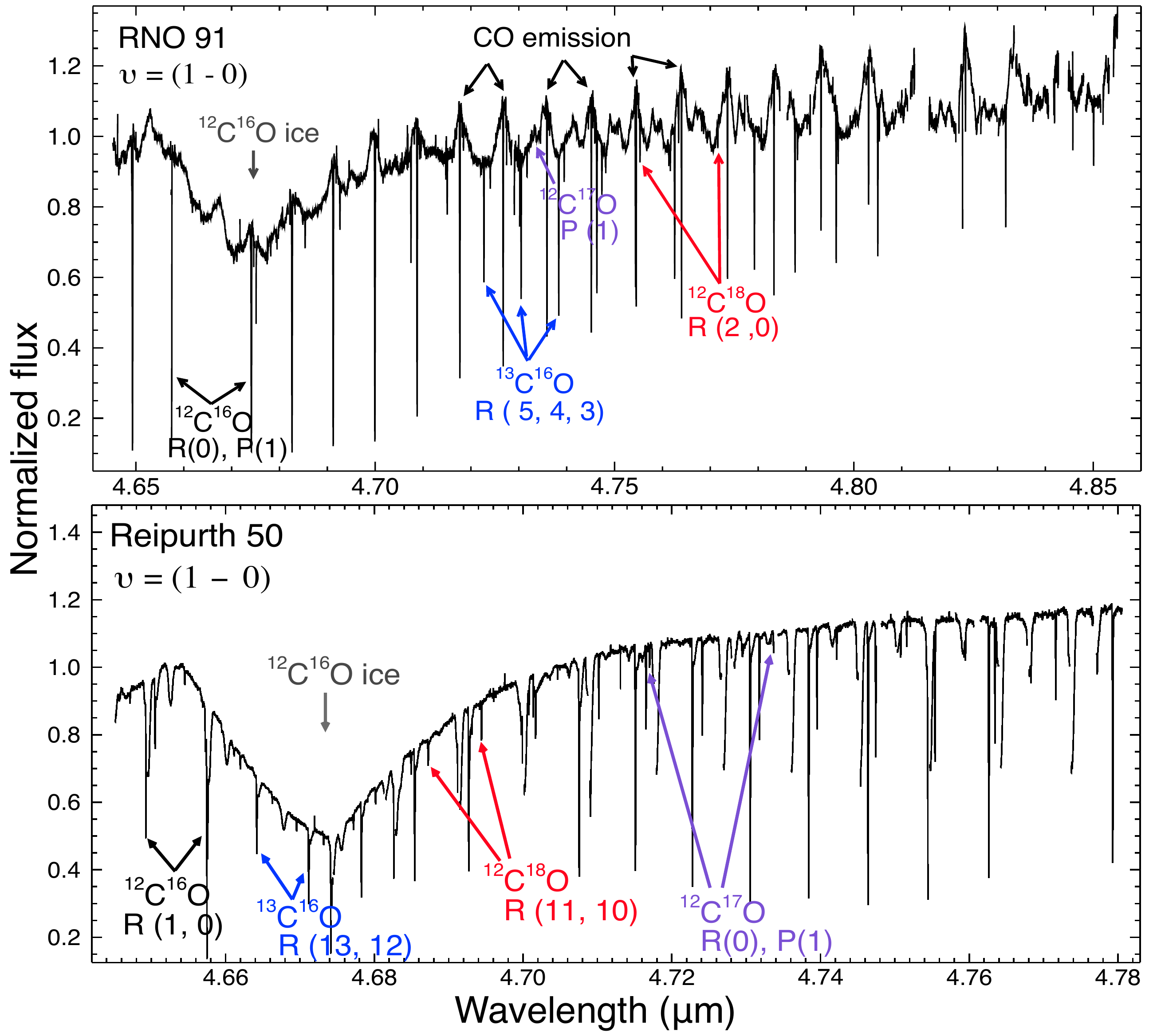}
\caption[]{Infrared absorption spectra of the CO fundamental rovibrational band toward RNO 91 and Reipurth 50, with moderate optical depths of CO ice $(\tau<$1).The rich CO forest of narrow rovibrational absorption lines are due to the various CO isotopologues present. Sample lines are marked. RNO 91 shows low-level emission possibly representative of the inner disk.}
\label{specm2a}
\end{figure}

\begin{figure}[ht]
\centering
\includegraphics[width=8.5cm]{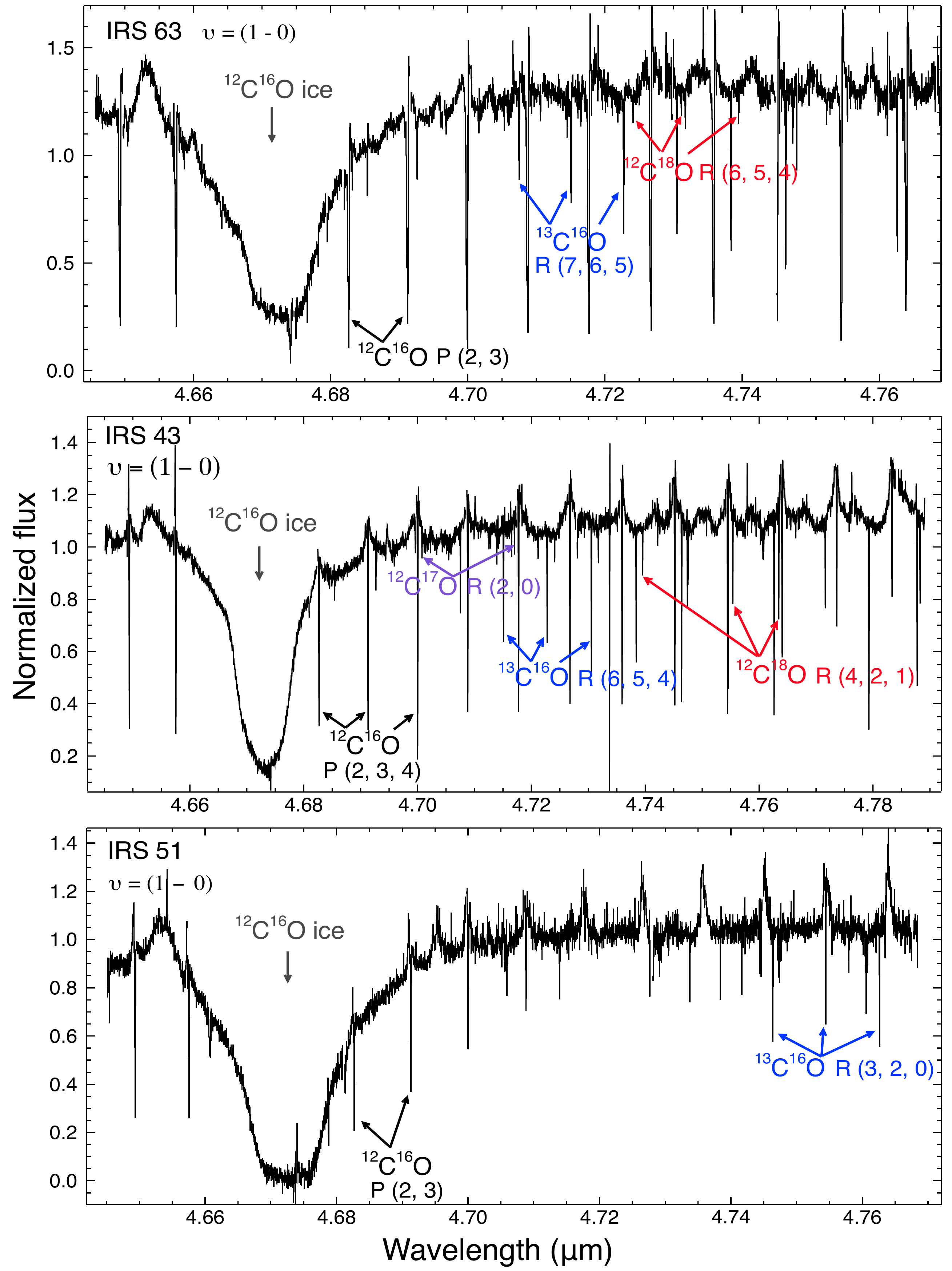}
\caption[]{Infrared absorption spectra of the CO fundamental rovibrational band toward IRS 63, IRS 43, and IRS 51, embedded YSOs with deep CO ice features. The CO absorption toward IRS 51 is particularly cold, leading to detection of only low-$J$ transitions. Sample lines are marked.}
\label{specm2b}
\end{figure}

\begin{figure*}[ht]
\centering
\includegraphics[width=18cm]{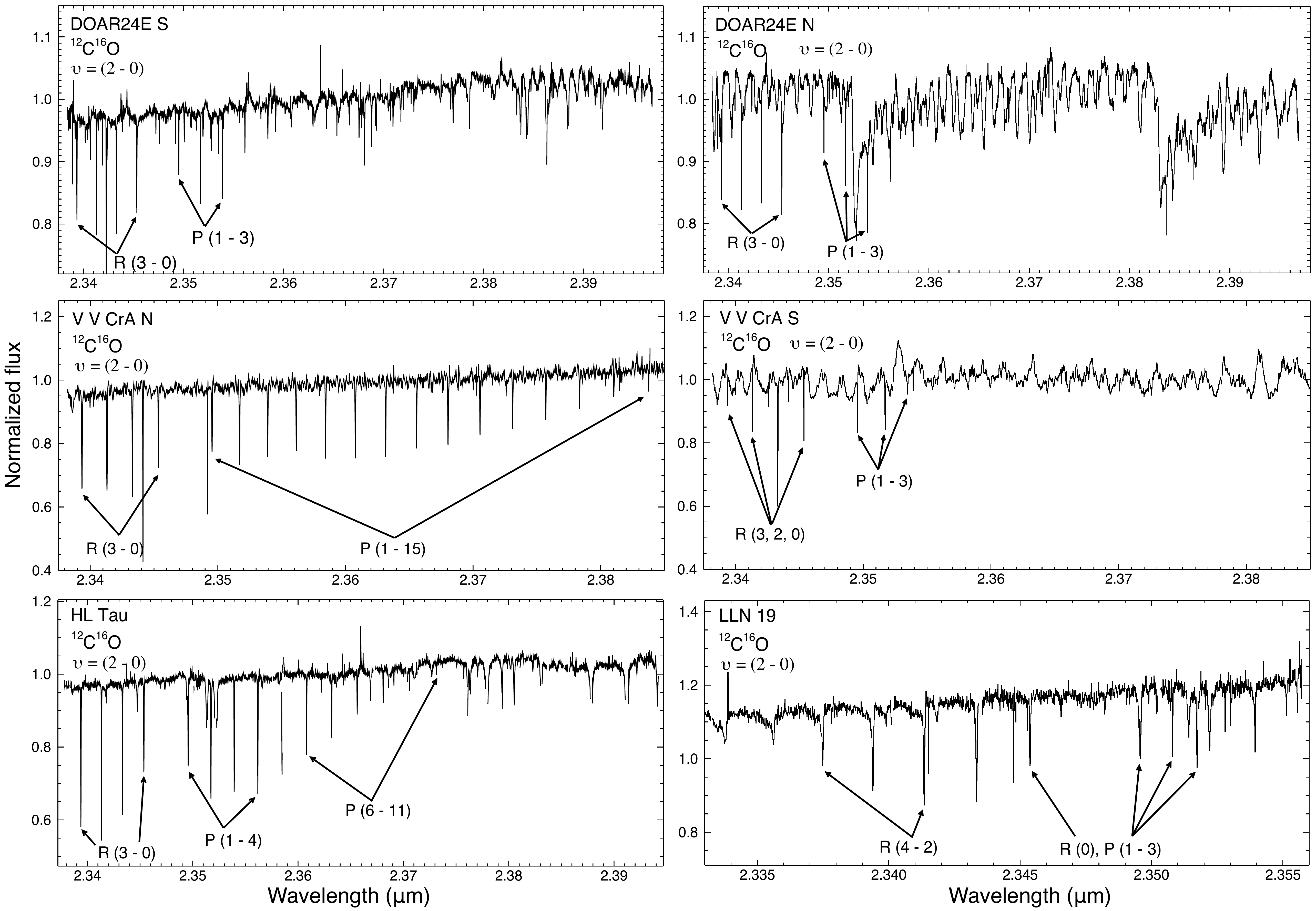}
\caption[]{Infrared absorption spectra of the CO rovibrational first-overtone bands toward the YSOs shown in Figures \ref{specm1a} and \ref{specm1b}. Several detected rovibrational $^{12}$C$^{16}$O lines are marked. Broad absorption lines at $\sim2.34-2.353\,\mu$m, most evident toward DoAR24E N,S, VV CrA S, and LLN 19, are due to photospheric CO.}
\label{speck1}
\end{figure*}

\begin{figure*}[ht]
\centering
\includegraphics[width=18cm]{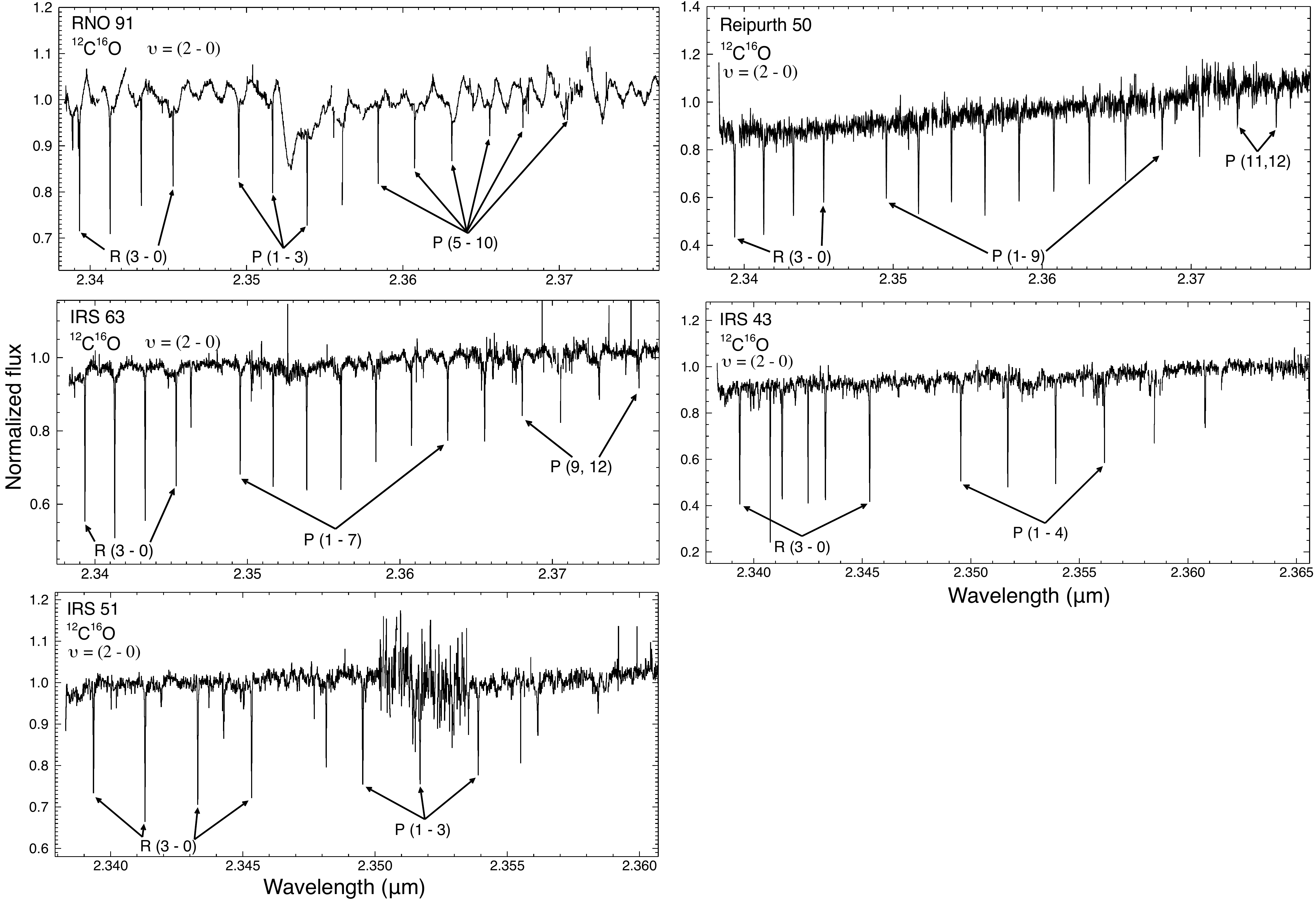}
\caption[]{Infrared absorption spectra of the CO rovibrational first-overtone bands toward the YSOs shown in Figures \ref{specm2a} and \ref{specm2b}. Several detected rovibrational $^{12}$C$^{16}$O lines are marked. Broad photospheric CO lines are observed toward RNO 91 and more weakly toward IRS 63.}
\label{speck2}
\end{figure*}

\begin{figure}[ht]
\centering
\includegraphics[width=8.5cm]{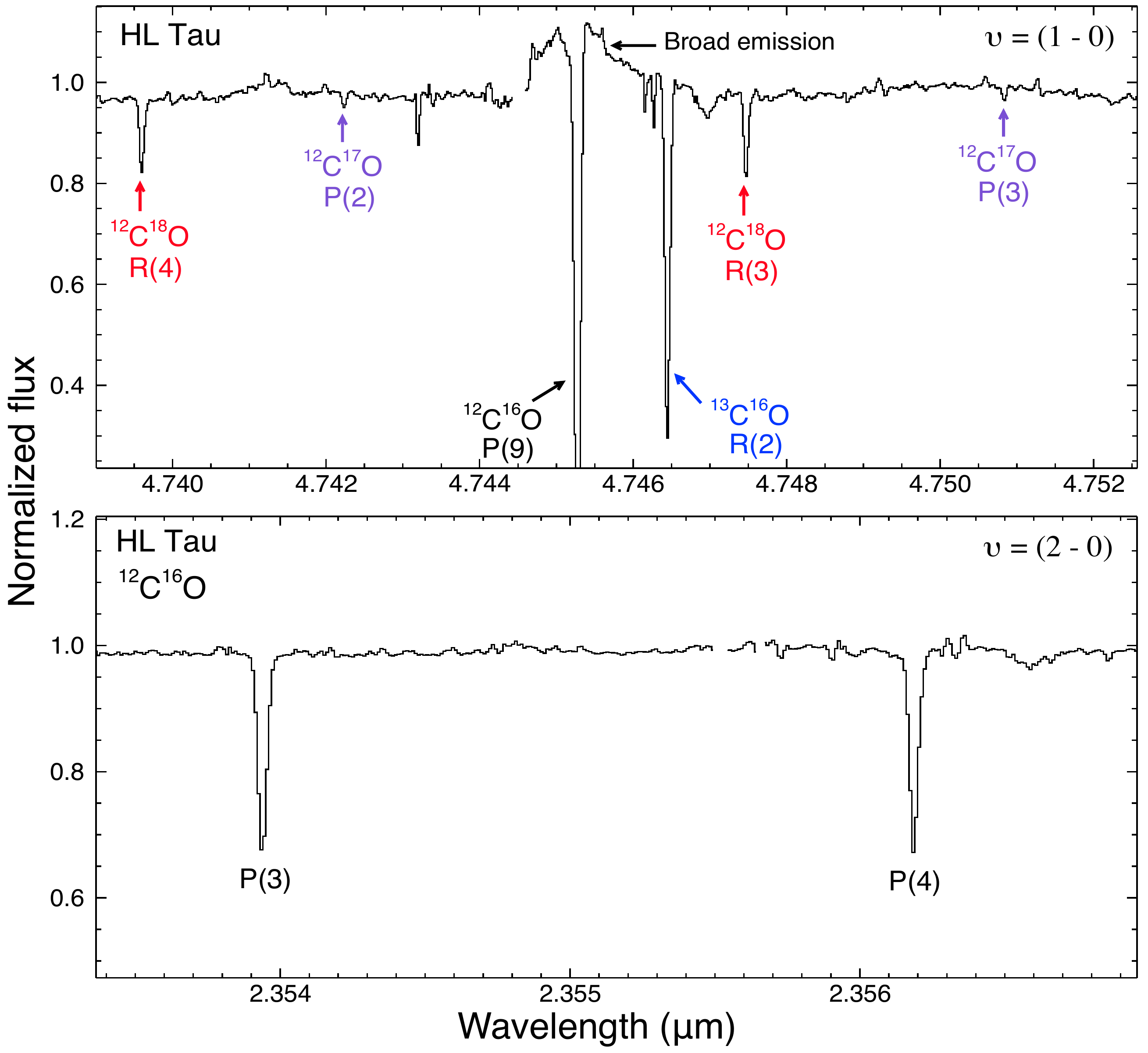}
\caption[]{Representative magnification for fundamental (top) and first-overtone (bottom) bands, here shown for HL Tau. Isotopologue lines are marked.}
\label{spec_zoom}
\end{figure}

\section{Methods and results}
\subsection{$\rm{CO}$ spectra} 
The fundamental $(v=1-0)$ and first-overtone $(v=2-0)$ rovibrational bands are shown for all targets in Figures \ref{specm1a} through \ref{speck2}, in ascending order of CO ice optical depths at $\sim4.675\,\mu$m. A forest of narrow CO rovibrational absorption lines is seen to varying degrees in all spectra. A magnified spectral region of HL Tau is shown as a representative example of individual line profiles (Figure \ref{spec_zoom}). Broad, complex $(v=1-0)$ CO emission is often seen in the spectra, and likely originates in the innermost ($R<1\,$AU) regions of the disks \citep{herczeg2011}. However, the emission does not interfere with the analysis, as it is much broader than the absorption, and is generally weak or absent in the rare isotopologue transitions. The potentially affected $(v=1-0)$ $^{12}$C$^{16}$O lines are not used for column density measurements. Broad absorption observed toward several objects at $(v=2-0)$ (Figures \ref{speck1}, \ref{speck2}) likely originates in the stellar photosphere and is clearly separated from the narrow interstellar CO lines.

Temporal variation in [$^{12}$CO]/[$^{13}$CO] was found toward VV CrA S, with a [$^{12}$CO]/[$^{13}$CO] abundance ratio of $136\pm1.3$ when using a $(v=1-0)$ observation four months prior to the first-overtone $(v=2-0)$, as compared to $127\pm1$ when coeval observations were used (errors are statistical 1$\sigma$ values). The possibility for temporal spectral variability should therefore be considered for DoAr24E and IRS 63, targets for which only non-coeval observations are available. 

\subsection{Derivation of column densities and isotopologue ratios}
\label{sec:derivation}
Total abundances for each isotopologue were derived by measuring the column density of CO in each observed rovibrational level of the ground vibrational state which, in turn, was obtained from the intrinsic optical depths at line center $(\tau _{\circ})$ derived from a best-fit absorption line profile to each rovibrational transition. Individual column densities were used to make a best fit to a Boltzmann population, then summed over all levels. This method is described in detail in \cite{smith2009}. 

\begin{deluxetable*}{llllllll}[ht]
\centering
\tablecolumns{8}
\tablewidth{0pt}
\tablecaption{Summary of observed line parameters}
\tablehead{
  \colhead{Target} & \multicolumn{2}{c}{FWHM$_{\rm instrumental}$\tablenotemark{a}}              & \colhead{FWHM$_{\rm intrinsic}$\tablenotemark{b}} &\colhead{$\tau(^{12}$C$^{16}$O)} &\colhead{$\tau(^{13}$C$^{16}$O)} &\colhead{$\tau(^{12}$C$^{18}$O)} &\colhead{$\tau(^{12}$C$^{17}$O)} \\
                   & \colhead{$M$ (km s$^{-1}$)} & \colhead{$K$ (km\,s$^{-1}$)} & \colhead{(km\,s$^{-1}$)}         &\colhead{$v=(2-0)$}  &\colhead{$v=(1-0)$} &\colhead{$v=(1-0)$} &\colhead{$v=(1-0)$} 
     }
\startdata
DoAr24E S   & 2.79 & 3.12 & $2.33\pm0.56$                   & $0.184-0.353$ & $0.217-1.033$ & $0.027-0.077$ & --           \\
DoAr24E N   & 2.79 & 3.12 & $2.33\pm0.56$\tablenotemark{c}  & $0.075-0.330$ & $0.198-1.014$ & $0.047-0.068$ & --           \\
VV CrA N    & 2.82 & 3.49 & $4.28\pm0.07$                   & $0.064-0.573$ & $0.568-1.322$ & $0.095-0.174$ &$0.020-0.038$ \\
VV CrA S    & 2.82 & 3.49 & $1.74\pm0.12$                   & $0.170-0.614$ & $0.225-1.377$ & $0.119-0.228$ &$0.021-0.034$ \\
HL Tau      & 3.26 & 3.43 & $2.63\pm0.05$                   & $0.027-0.987$ & $0.036-2.896$ & $0.079-0.310$ &$0.021-0.052$ \\
LLN 19       & 3.22 & 4.68 & $3.52\pm0.27$                   & $0.196-0.437$ & $0.627-0.909$ & $0.072-0.108$ & --           \\
RNO 91      & 3.26 & 3.72 & $1.67\pm0.27$                   & $0.105-0.843$ & $0.089-2.290$ & $0.046-0.370$ &$0.059-0.072$ \\
Reipurth 50 & 3.44 & 4.54 & $3.08\pm0.02$                   & $0.225-1.366$ & $0.338-2.831$ & $0.027-0.707$ &$0.044-0.138$ \\
IRS 63      & 3.57 & 3.39 & $3.83\pm0.35$                   & $0.057-0.811$ & $0.171-1.582$ & $0.141-0.245$ &$0.043-0.081$ \\
IRS 43      & 3.03 & 3.45 & $1.86\pm0.11$                   & $1.076-2.563$ & $0.212-4.542$ & $0.073-1.115$ &$0.033-0.280$ \\
IRS 51      & 3.03 & 3.37 & $1.88$\tablenotemark{d}         & $0.504-0.822$ & $0.739-1.437$ & --            & -- 
\enddata
\tablecomments{
\tablenotetext{a}{The instrumental broadening is measured by fitting an atmospheric model to unresolved telluric lines. The uncertainty is $\lesssim0.1$ km s$^{-1}$. These refined measurements of instrumental broadening account for the slight variation in results for VV CrA and Reipurth 50 as compared to \citep{smith2009}.}
\tablenotetext{b}{The intrinsic velocity dispersion is the Gaussian FWHM before being convolved with the measured instrumental CRIRES broadening for the individual spectra. It is based on a weighted mean FWHM from a selection of $^{12}$C$^{18}$O lines, unless otherwise noted. Uncertainties reflect the weighted means of the individual $^{12}$C$^{18}$O line widths.} 
\tablenotetext{c}{Mean value from DoAr24E S $^{12}$C$^{18}$O lines $R(2)$ and $R(1)$.}
\tablenotetext{d}{This width is estimated based $^{12}$C$^{16}$O and $^{13}$C$^{16}$O lines, since the source has very weak $^{12}$C$^{18}$O lines.}
}
\label{tau_table}
\end{deluxetable*}

The technique of using the line profiles to directly determine column densities takes advantage of the high spectral resolution of CRIRES to spectrally resolve the velocity dispersion of the individual CO lines. This in turn leads to greater precision than using unresolved lines which require a curve-of-growth analysis to derive intrinsic line widths \citep{mitchell1989, mitchell1990, mitchell1991, sheffer1992, mitchell1993, lambert1994, sheffer2002, federman2003, goto2003, brittain2005}. With CRIRES, the resolving power is sufficiently high to partially resolve the intrinsic line profiles (FWHM$_{\rm Intrinsic} \sim{\rm FWHM}_{\rm Instrumental}$), albeit marginally in some cases: VV CrA S, RNO 91, and IRS 43. Therefore, it is possible that the derived column densities toward these targets may be more uncertain. 

\begin{deluxetable}{llll}[ht]
\centering
\tablecolumns{4}
\tablewidth{0pt}
\tablecaption{Optical depths ($\tau_{\circ}$) and Doppler shifts for CO isotopologue rovibrational lines used in deriving column densities for DoAr24E S\tablenotemark{a,}\tablenotemark{b}}

\tablehead{
\colhead{Isotopologue}  & \colhead{Transition} & \colhead{$\tau_0$} & \colhead{$V_{\rm LSR}$} \\
                       &                      &                           & {(km s$^{-1}$)} 
     }
\startdata
$^{12}$C$^{16}$O  & {$(2, 0)\ R(3)$}   & $0.279\pm0.002$ & $3.09\pm0.01$ \\
$^{12}$C$^{16}$O  & {$(2, 0)\ R(2)$}   & $0.353\pm0.002$ & $3.31\pm0.01$ \\
$^{12}$C$^{16}$O  & {$(2, 0)\ R(1)$}   & $0.326\pm0.002$ & $3.42\pm0.01$ \\
$^{12}$C$^{16}$O  & {$(2, 0)\ R(0)$}   & $0.262\pm0.002$ & $3.84\pm0.02$ \\
$^{12}$C$^{16}$O  & {$(2, 0)\ P(1)$}   & $0.184\pm0.002$ & $3.53\pm0.02$ \\
$^{12}$C$^{16}$O  & {$(2, 0)\ P(2)$}   & $0.255\pm0.002$ & $3.47\pm0.01$ \\
$^{12}$C$^{16}$O  & {$(2, 0)\ P(3)$}   & $0.250\pm0.003$ & $4.06\pm0.03$ 
\enddata
\tablecomments{
\tablenotetext{a}{This is a partial table for DoAr24E S. Line-by-line tables for all observed targets are available in the online version of the paper, and can be temporarily found at {\tt http://naturalsciences.org/data/apj2015}.}
\tablenotetext{b}{Uncertainties are $1\sigma$ derived from the line fits.}
}
\label{line_tab}
\end{deluxetable}

The measured instrumental broadening, intrinsic velocity dispersions and optical depth ranges for each observed isotopologue are shown in Table \ref{tau_table}. Line-by-line optical depths and Doppler shifts for observed targets are available in the online version of this paper, with a sample partial line list shown in Table \ref{line_tab}. Complete data tables are available in the online version of the paper\footnote{Complete data tables can be temporarily found at {\tt http://naturalsciences.org/data/apj2015}.}. Instrumental broadening values were modeled individually for each spectrum in its entirety, following the method in \cite{Seifahrt2010} for using an atmospheric model for telluric spectra. The Reference Forward Model (RFM; http://www.atm.ox.ac.uk/RFM/) code was used to derive a detailed fit to the individual telluric standard star spectra, with consideration of meteorological conditions at the time of observation. The atmospheric molecular parameters were taken from the HITRAN 2008 database \citep{rothman2009}. There are small relative differences in resolving power between the standard stars and science targets due to differences in seeing or adaptive optics (AO) correction, but these differences were accurately measured and corrected for by minimizing the continuum residuals after telluric division. 

Optimized intrinsic velocity dispersions were derived from the weighted mean of dispersions measured in individual $^{12}$C$^{18}$O lines, assuming that both the intrinsic and instrumental profiles are well-approximated by Gaussian profiles. The $^{12}$C$^{18}$O lines were used because this rare isotopologue is always optically thin, and therefore is not subjected to saturation line broadening, which may affect the more common $^{13}$CO isotopologue. Optical depths of the optically-thin $^{12}$C$^{16}$O $(v=2-0)$ lines were similarly computed using the derived intrinsic broadening from the $(v=1-0)$ $^{12}$C$^{18}$O lines. It can be noted that, even when the resolving power of the two spectral bands differed for a single target, we obtained good fits to lines in both bands under the reasonable assumption that the intrinsic profile width is independent of transition and isotopologue. Further, it is reasonable to use the same velocity dispersion parameter for a target with a two-temperature distribution of the data, since thermal contributions are very small contributors to the total intrinsic dispersion (0.077 and 0.242 km s$^{-1}$ for $^{12}$C$^{16}$O at 20 and 200 K, respectively).

\begin{figure*}[ht]
\centering
\includegraphics[width=16cm]{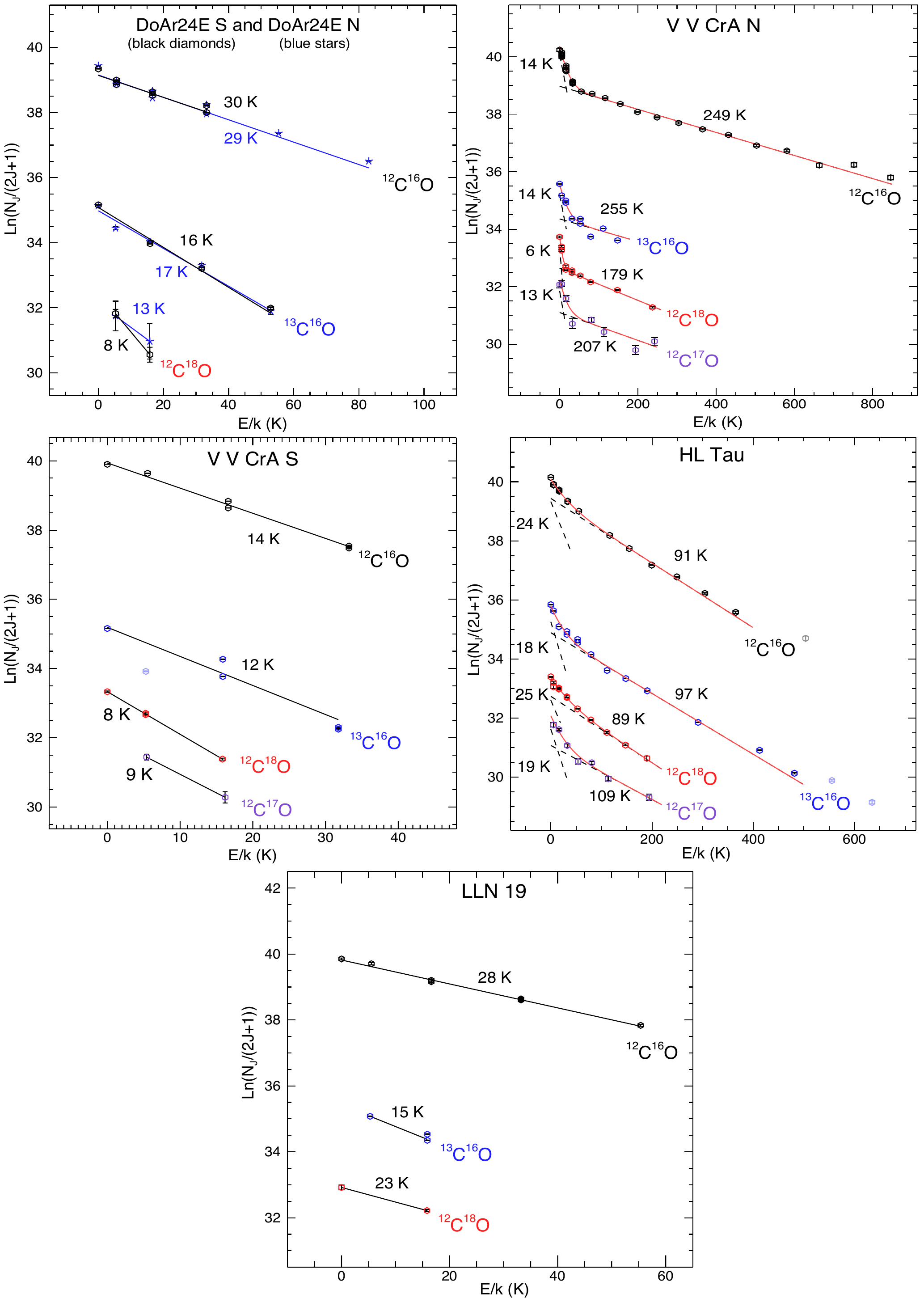}
\caption[]{Rotational excitation diagrams for targets with little to no ice, corresponding to targets shown in Figures \ref{specm1a}, \ref{specm1b}, and \ref{speck1}). $E$ is the energy of the $J$th rotational state above the ground rotational state and $k$ is the Boltzmann constant. One- and two-temperature simultaneous fits were used as warranted by the rotational excitation structure. Derived high and low temperatures for each isotopologue are indicated on the plots. Error bars are $1\sigma$ propagated from the Gaussian fits to the line profiles. Faded symbols are outliers that have been excluded from the fits.}
\label{roplots1}
\end{figure*}

\begin{figure*}[ht]
\centering
\includegraphics[width=16cm]{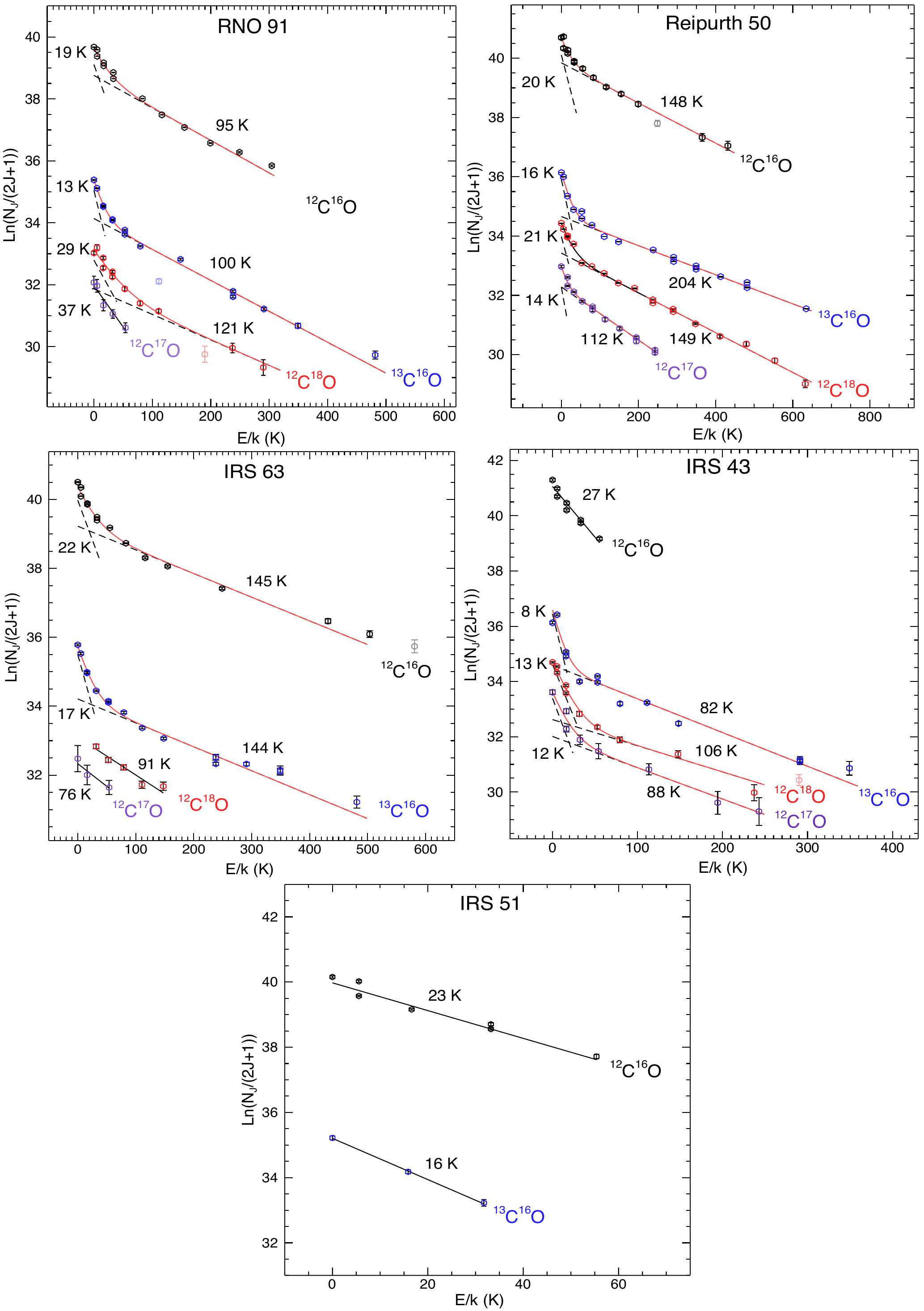}
\caption[]{Rotational excitation diagrams for targets with moderate to deep CO ice features, corresponding to targets in Figures \ref{specm2a}, \ref{specm2b} and \ref{speck2}. $E$ is the energy of the $J$th rotational state above the ground rotational state and $k$ is the Boltzmann constant. One- and two-temperature simultaneous fits were used as warranted by the rotational excitation structure. Derived high and low temperatures for each isotopologue are indicated on the plots. Error bars are $1\sigma$ propagated from the Gaussian fits to the line profiles. Faded symbols are outliers that have been excluded from the fits.}
\label{roplots2}
\end{figure*}

The rotational analyses for all targets are shown in Figures \ref{roplots1} and \ref{roplots2}. Sub-level column densities, $N_{J}/(2J + 1)$, derived from the fits to the absorption lines, are plotted against the energy, $E/k$, of the rotational transition state, $J$. For targets where cold gas dominates, a single-temperature model is sufficient to fit the data, while targets with absorption that includes higher-$J$ transitions are best fit with a two-temperature model, representing cold and warm environments along the line-of-sight, and allowing for different excitation temperatures for each isotopologue along a given line-of-sight. Derived isotopologue temperatures from the fits are shown in Table \ref{temp_table}. High temperatures ($\sim80$ to 250 K) are roughly consistent with those of the inner parts of typical YSO envelopes, while low temperatures ($\sim5$ to 30 K) may trace outer envelopes or molecular cloud material in front of the YSOs. For lines-of-sight with no warm component, it is likely that the gas is from a foreground cloud not related to the YSO itself (DoAr24E N and S), VV CrA S, and IRS 51. 

Total column densities and isotopologue ratios are shown in Table \ref{ratio_table}. Due to the dominance of higher-$J$ lines in targets with both low- and high-temperature lines, the higher-temperature ratios are generally better constrained within two-temperature environments \citep{smith2009}. 

Statistical errors on the column densities are generally found to be small, reflecting the very high signal-to-noise ratio of the data. For example, average signal-to-noise per pixel in the fundamental bands is $\sim250$. With an average CRIRES FWHM of 2.0 pixels, we retrieve average line errors of $\sim0.3\%$ (1/($250\times\sqrt{2.0}$)). First-overtone bands average $\sim1\%$. We therefore expect to know the column densities to a fraction of a percent, which is represented in our results. These errors do not take into account any potential systematic errors in extracting the line strengths and profiles due to inherent non-gaussianity of the line-fitting procedure, which we cannot quantify with certainty. However, to roughly estimate the magnitude of systematic retrieval errors, we carried out a small number of forward models with 2-dimensional radiative transfer models \citep[RADLite, ][]{Pontoppidan2009} simulating synthetic CO isotopologue absorption spectra using a basic one-dimensional infalling envelope from \cite{shu77}. The synthetic spectra were convolved with the appropriate instrumental resolution and isotopologue ratios retrieved from the synthetic spectra using the same methods as for the observed data. In each case, we were able to reproduce the input ratios to better than 10\%. We therefore estimate that model-dependent systematics add at most a 10\% error to the retrieved isotopologue ratios, in addition to the statistical error.

\begin{deluxetable*}{llcccc}[ht]
\centering
\tablecolumns{6}
\tablewidth{0pt}
\tablecaption{Measured CO isotopologue rotational temperatures}

\tablehead{
\colhead{Target} & \colhead{Component\tablenotemark{a}} & \colhead{$T_{\rm ex}(^{12}$C$^{16}$O)}  &\colhead{$T_{\rm ex}(^{13}$C$^{16}$O)}  &\colhead{$T_{\rm ex}(^{12}$C$^{18}$O)}  &\colhead{$T_{\rm ex}(^{12}$C$^{17}$O)}\\
                 &                     & (K) & (K) & (K) & (K)
     }
\startdata 
DoAr24E S   & single & $30\pm 1$  & $16\pm 1$  & $8\pm 2$    & --          \\
DoAr24E N   & single & $29\pm 1$  & $17\pm 1$  & $13\pm 12$  & --          \\
VV CrA N    & high   & $249\pm 2$ & $255\pm 8$ & $179\pm 5$  & $207\pm 39$ \\
VV CrA N    & low    & $14\pm 1$  & $14\pm 1$  & $6\pm 1$    & $13\pm 3$   \\  
VV CrA S    & single & $14\pm 1$  & $12\pm 1$  & $8\pm 1$    & $9\pm 1$    \\
HL Tau      & high   & $91\pm 1$  & $97\pm 1$  & $89\pm 6$   & $109\pm 17$ \\
HL Tau      & low    & $24\pm 1$  & $18\pm 1$  & $25\pm 3$   & $19\pm 5$   \\
LLN 19       & single & $28\pm 1$  & $15\pm 1$  & $23\pm 2$   & --          \\
RNO 91      & high   & $95\pm 1$  & $100\pm 1$ & $121\pm 21$ & --          \\
RNO 91      & low    & $19\pm 1$  & $13\pm 1$  & $29\pm 5$   & $37\pm 6$   \\
Reipurth 50 & high   & $148\pm 8$ & $204\pm 1$ & $149\pm 1$  & $112\pm 3$  \\
Reipurth 50 & low    & $20\pm 3$  & $16\pm 1$  & $21\pm 1$   & $14\pm 1$   \\
IRS 63      & high   & $145\pm 3$ & $144\pm 4$ & $91\pm 8$   & $76\pm 40$  \\
IRS 63      & low    & $22\pm 1$  & $17\pm 1$  & --          & --          \\
IRS 43      & high   & $27\pm1$   & $82\pm 2$  & $106\pm15$  & $88\pm21$   \\
IRS 43      & low    & --         & $8\pm1$    & $13\pm 1$   & $12\pm 2$   \\
IRS 51      & single & $23\pm 1$  & $16\pm 1$  & --          & --          

\enddata
\tablecomments{
\tablenotetext{}{All uncertainties are $1\sigma$ statistical error derived from the rotational analysis.}
\tablenotetext{a}{The temperature component. Single if a single temperature component was used. High and low for two-component fits.}
}
\label{temp_table}
\end{deluxetable*}

\subsection{Notes on individual sources}
The DoAr24E system is a hierarchical multiple for which we have spectra of two components, with the rotational analyses nearly identical for both components. This lack of significant spatial variability strongly suggests that we are tracing spatially extensive and homogenous foreground material, that in turn is the dominant contributor smoothly distributed on scales of at least a few 100 AU. The lack of isotopic variability between the DoAr24E components gives us further confidence on our retrieval method for deriving precise isotope ratios. 

Due to high extinction leading to a faint $v=(2-0)$ band continuum for the embedded YSO IRS 43, we detect high-$J$ lines for all CO isotopologues in $v=(1-0)$, but only low-$J$ lines for $^{12}$C$^{16}$O in the $v=(2-0)$ band. Given this disparate set of temperature ranges, our data for IRS 43 are therefore not useful for measuring CO isotopologue abundance ratios relative to $^{12}$C$^{16}$O. 

Toward VV CrA S, we derive a $^{12}$C$^{16}$O column density of $1.178\pm0.004\times10^{18}$\,cm$^{-2}$. This is nearly three orders of magnitude higher than the value of $2.6\pm0.4\times10^{15}$\,cm$^{-2}$ derived in \cite{kruger2011}. Our rotational temperature ($14\pm1$\,K) is also significantly lower than their value ($29\pm4$\,K). Since \cite{kruger2011} used the $(v=1-0)$ spectra from the archive, these differences could be explained by strong saturation effects resulting from their use of the highly optically thick $v=(1-0)$ $^{12}$C$^{16}$O lines. This difference reinforces the necessity of using $v=(2-0)$ data for deriving precise $^{12}$C$^{16}$O column densities. 

\cite{smith2009} proposed that the cold gas component observed toward VV CrA S may be interpreted as either molecular cloud material in front of the VV CrA binary system, or disk gas toward the primary component as suggested by \cite{kruger2011}. We point out that the former is more likely because the presence of both warm and cold components are expected for a disk, while a foreground cloud will typically only show a single cold component, such as that seen towards VV CrA N. Taken together, the likely geometry for the VV CrA system and the spectroscopic results render it unlikely that the VV CrA S spectra represent outer disk. VV CrA N observations most likely represent warm and cold disk gas. 

\begin{deluxetable*}{llccccc}[ht]
\centering
\tablecolumns{7}
\tablewidth{0pt}
\tablecaption{Derived $^{12}$C$^{16}$O Isotopologue Column Densities and Isotopologue Ratios}

\tablehead{
\colhead{Target} & \colhead{Comp.} & \colhead{$N(^{12}$C$^{16}$O)}  & \colhead{[$^{12}$C$^{16}$O]/[$^{13}$C$^{16}$O]} & \colhead{[$^{12}$C$^{16}$O]/[$^{12}$C$^{18}$O]}  & \colhead{[$^{12}$C$^{16}$O]/[$^{12}$C$^{17}$O]}  & \colhead{[$^{12}$C$^{18}$O]/[$^{12}$C$^{17}$O]}\\
	& &  \colhead{($10^{18}$ cm$^{-2}$)} & & & &
     }
\startdata 
DoAr24E S                  & single & $1.098\pm 0.005$ & $98\pm 1$    & :$2500\pm 560$   & --             & -- \\
DoAr24E N                  & single & $1.094\pm 0.003$ & $102\pm 3$   & :$2200\pm 1620$  & --             & -- \\
VV CrA N                   & high   & $7.618\pm 0.074$ & $95\pm 1$    & $740\pm 15$     & $3085\pm 440$  & $4.1\pm 0.6$ \\
VV CrA N                   & low    & $1.413\pm 0.023$ & $127\pm 2$   & $1750\pm 70$    & $4000\pm 630$  & $2.3\pm 0.4$ \\
VV CrA S                   & single & $1.178\pm 0.004$ & $127\pm 1$   & :$1150\pm 20$    & :$3850\pm 580$  & $3.4\pm 0.5$ \\
HL Tau                     & high   & $4.467\pm 0.064$ & $85\pm 1$    & $795\pm 80$     & $3500\pm 660$  & $4.4\pm 0.9$ \\
HL Tau                     & low    & $1.067\pm 0.016$ & $74\pm 1$    & $760\pm 80$     & $2670\pm 370$  & $3.5\pm 0.6$ \\
LLN 19                     & single & $2.020\pm 0.034$ & $137\pm 3$   & :$1150\pm 90$    & --             & -- \\
RNO 91\tablenotemark{a}    & high   & $2.364\pm 0.032$ & $92\pm 2$    & $740\pm 210$    & $2170\pm 280$  & $2.9\pm 0.9$ \\
RNO 91\tablenotemark{a}    & low    & $0.689\pm 0.007$ & $80\pm 2$    & $350\pm 40$     & $630\pm 80$    & $1.8\pm 0.3$ \\
Reipurth 50                & high   & $10.81\pm 0.766$ & $124\pm 9$   & $580\pm 40$     & $2480\pm 200$  & $4.3\pm 0.2$ \\
Reipurth 50                & low    & $1.976\pm 0.157$ & $76\pm 6$    & $420\pm 30$     & $3470\pm 380$  & $8.2\pm 0.6$ \\
IRS 63                     & high   & $5.743\pm 0.151$ & $145\pm 6$   & $690\pm 60$     & $1830\pm 510$  & $2.6\pm 0.8$ \\
IRS 63                     & low    & $1.904\pm 0.026$ & $110\pm 3$   & --              & --             & -- \\
IRS 43\tablenotemark{b}    & high   & $6.908\pm 0.158$ & :$210\pm 10$ & :$1160\pm 180$  & :$2600\pm 960$ & $2.2\pm 0.9$ \\
IRS 43\tablenotemark{b}    & low    & --               & :$300\pm 10$ & :$1200\pm 40$   & :$4500\pm 580$ & :$3.7\pm 0.5$ \\
IRS 51                     & single & $2.026\pm 0.036$ & $165\pm 10$  & --              & --             & -- \\
Local ISM\tablenotemark{c} & --     & --               & $69\pm 6$    & $557\pm 30$     & $2005\pm 155$  & $3.6\pm 0.2$ 

\enddata
\tablecomments{
\tablenotetext{}{All uncertainties are $1\sigma$ statistical errors derived from the line fits.}
\tablenotetext{}{Isotope ratios that may be affected by significant systematic uncertainty are indicated with a ``:''.}
\tablenotetext{a}{The $^{12}$C$^{17}$O lines for RNO 91 were fit with a single-temperature model.} 
\tablenotetext{b}{Ratios relative to $^{12}$C$^{16}$O are uncertain because only low-$J$ $^{12}$C$^{16}$O lines are available, while the low-$J$ lines for the other isotopologues are poorly constrained.} 
\tablenotetext{c}{Values from \cite{Wilson1999}.}
}
\label{ratio_table}
\end{deluxetable*}

\subsection{Carbon isotope ratios versus Galactocentric radius}
\begin{figure}[ht]
\centering
\includegraphics[width=8.5cm]{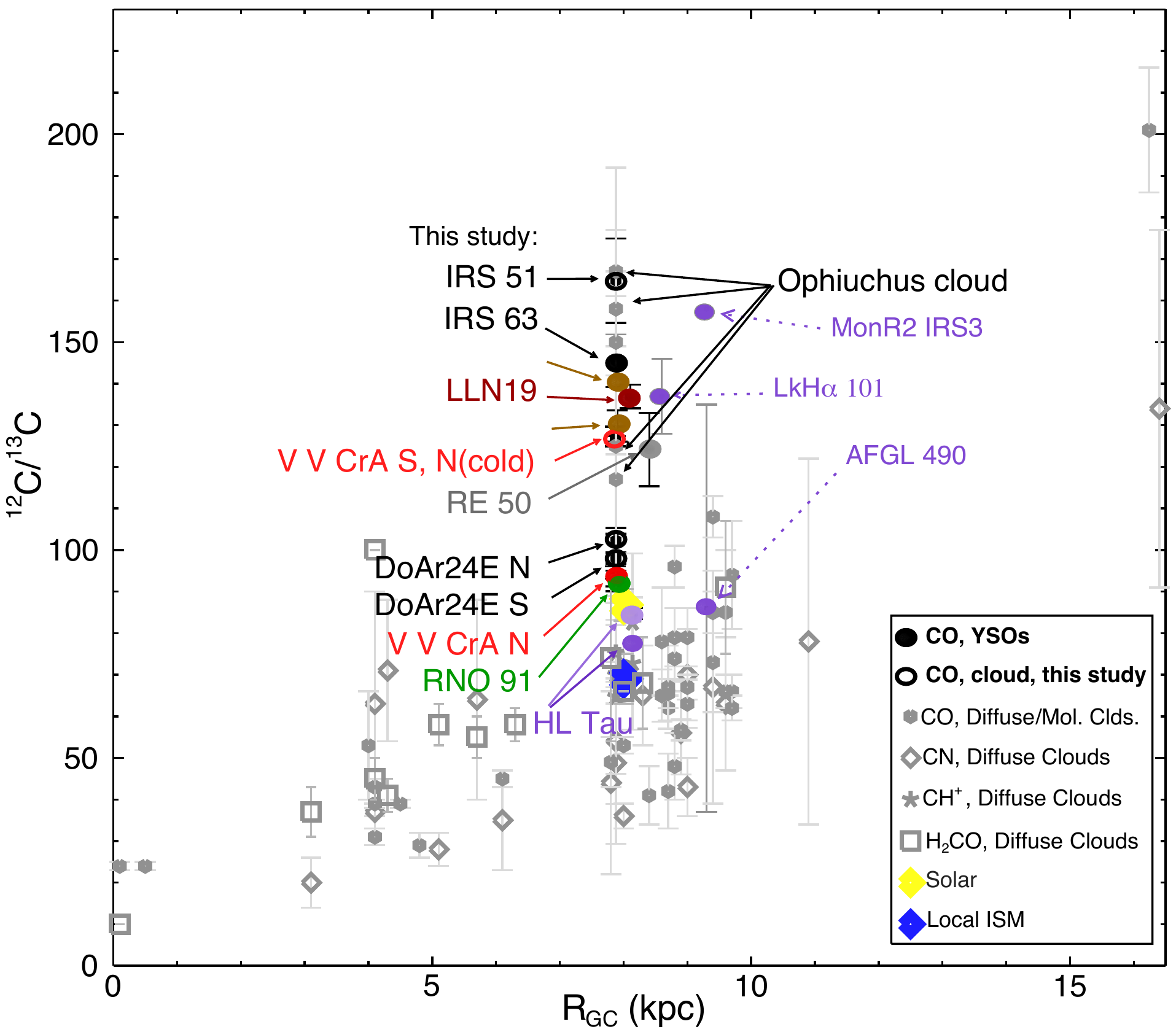}
\caption[]{Compilation of $^{12}$C/$^{13}$C ratios vs. R$_\textrm{\tiny GC}$ (kpc). Gray symbols show $^{12}$C/$^{13}$C values from the literature, derived from different carbon-bearing molecules, CO, CN, CH$^{+}$ and H$_{2}$CO. Data from our study are labeled for YSO targets (filled circles) and foreground clouds (unfilled circles). Colors connect these data by molecular cloud with those in Figure \ref{delta_delta}. For targets where there is both warm and cold gas along a line-of-site, ratios are shown for the more robustly determined warm gas, unless otherwise noted. Local ISM ($68\pm15$; \cite{Milam2005}) and solar system values ($86.8\pm 3.8$; \cite{scott2006}) are noted. The black arrows indicate high $^{12}$C/$^{13}$C in diffuse regions of Ophiuchus \citep{lambert1994, federman2003}. YSO data for MonR2 IRS3, LkH $\alpha$101, AFGL 490 \citep{goto2003} and HL Tau (lower value; \cite{brittain2005}) are marked (dark purple).}
\label{gc_co_plot}  
\end{figure}

Derived [$^{12}$C$^{16}$O]/[$^{13}$C$^{16}$O] ratios are plotted against Galactocentric radius (R$_\textrm{\tiny GC}$) in Figure \ref{gc_co_plot}. The solar system and local ISM values of $\sim87$ \citep{scott2006} and $\sim68$ \citep{Milam2005} are plotted at 8 kpc, respectively. Derived  [$^{12}$C$^{16}$O]/[$^{13}$C$^{16}$O] toward our targets range from $\sim85$ to 165, values inconsistent with those toward most YSOs and molecular clouds, or from GCE \citep{Prantzos1996} and regression models \citep{Milam2005}. The distribution of data from the literature approximates the general trend of increasing metallicity toward the Galactic center, with the corresponding increase in $^{12}$C/$^{13}$C with greater R$_\textrm{\tiny GC}$. 

\subsection{Derivation of $\rm{CO}$ ice fractions}
Previous studies have discussed the exchange between CO solid- and gas-phase reservoirs in disks and envelopes \citep[e.g.][]{vandishoeck1996, boogert2000}. With a range of CO ice column densities centered at $4.75\,\mu$m (Figures \ref{specm1a} through \ref{specm2b}), our data set provided an opportunity to investigate relationships between the gas-phase isotopologue ratios and the CO ice fraction along single lines-of-sight. CO ice column densities were calculated from optical depth analyses in \cite{Pontoppidan2003} and \cite {thi2010}.

Of the broad CO ice profile, the pure, cold CO component is most relevant with respect to interactions with CO gas, as it is the solid state reservoir that would most likely directly exchange with the gas in cold, dense clouds. We used optical depths for this component to derive total CO ice column densities for each target. Using the phenomenological decomposition analyses in \cite{Pontoppidan2003}, the pure component corresponds to the middle optical depth ($\tau_{\rm mc}$) profile measurement; \cite{thi2010} was used for LLN 19. For targets with no (VV CrA, DoAr24E) or very weak (HL Tau) spectral ice features, we used upper limits from the CRIRES fundamental bands.

Total column densities of pure CO in the ice phase $(N_{\rm Pure,CDE})$ were computed using the relation from \cite{Pontoppidan2003}, 
\begin{equation}
N_{\rm Pure, CDE} = 6.03 \, {\rm cm}^{-1} \times \tau_{\rm max} \times A_{\rm bulk}^{-1},
\label{eq:ice}
\end{equation}

\noindent
where $\tau_{\rm max}$ is the optical depth at line center of the pure CO component, $A_{\rm bulk}$ is the band strength of $^{12}$CO, and the numerical factor, 6.03\,cm$^{-1}$, considers the effect of grain shape on the band strength. In this study, we used the continuous distribution of ellipsoids (CDE) model \citep{Pontoppidan2003} in computing solid CO column densities.  

Optical depths, column densities, and total CO ice fractions are listed in Table \ref{ice_table}, in ascending order of CO ice optical depth. Although it is assumed that cold CO gas will more likely exchange with CO ice, for targets where CO gas was observed in both cold and warm environments, values for CO ice fractions are shown separately for each environment.

\begin{figure}
\centering
\includegraphics[width=8.5cm]{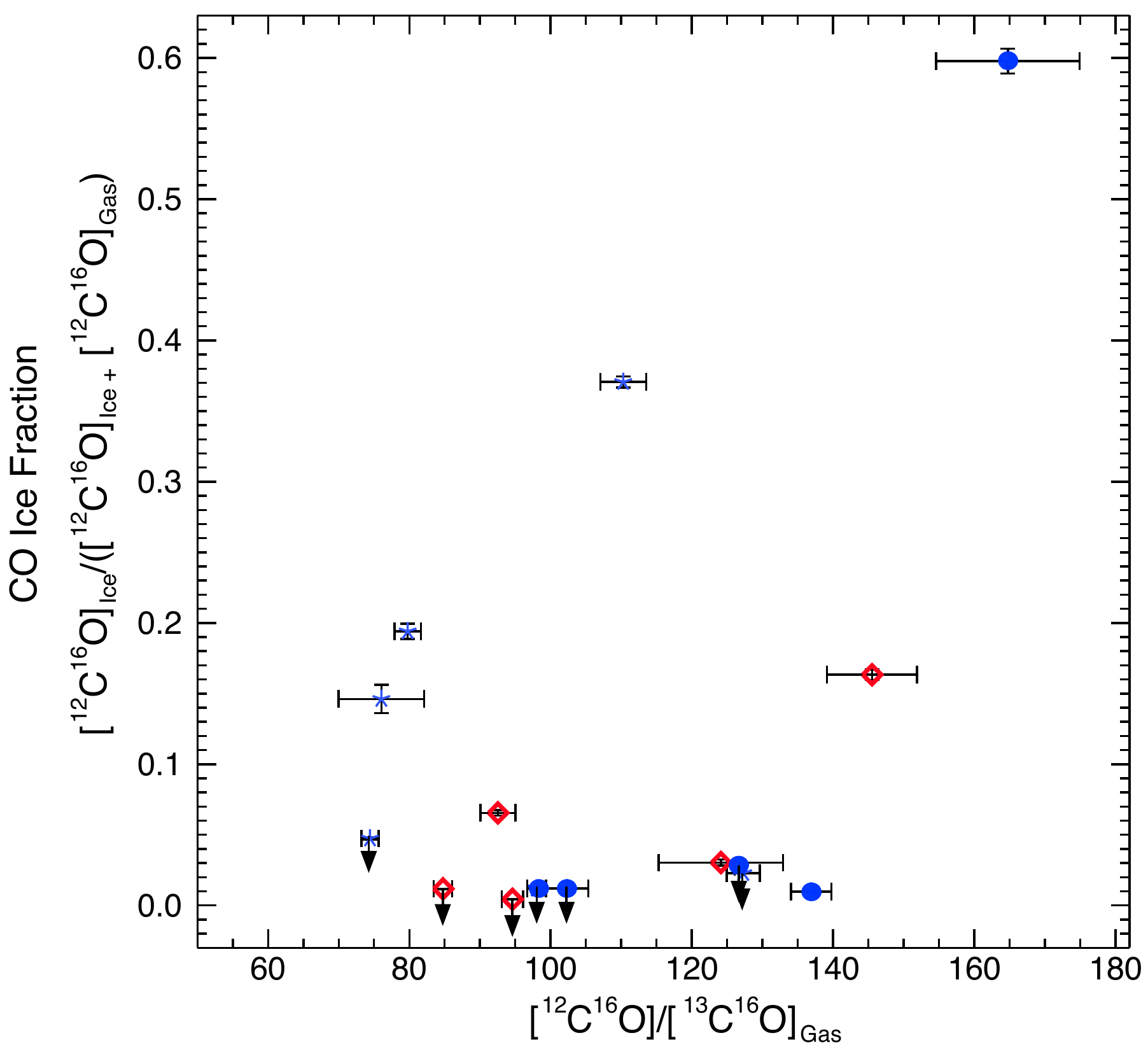}
\caption[]{Comparison of the total $^{12}$C$^{16}$O ice fraction with the [$^{12}$C$^{16}$O]/[$^{13}$C$^{16}$O] gas-phase ratios for targets observed in this study. For targets where both warm and cold gas was observed, CO ice fractions are plotted with red diamonds when warm gas was used in computing the fraction, and blue stars when cold gas was used. For targets where only cold gas was observed, CO ice fractions are shown with over-plotted blue ovals. Errors are $1\sigma$ propagated from errors in ice and gas optical depths measurements. $3\sigma$ upper limits (down-facing arrows) are shown for DoAr24E N,S, VV CrA N,S, and HL Tau.}
\label{ice_vs_ratio}
\end{figure}

The relation between the CO ice fraction and [$^{12}$C$^{16}$O]/[$^{13}$C$^{16}$O] gas-phase ratios is shown in Figure \ref{ice_vs_ratio}, with CO ice fractions plotted against derived [$^{12}$CO]/[$^{13}$CO] for cold ($T\sim5$ to 30\,K) and warm ($T\sim80$ to 250\,K) regimes, the latter only for targets where warm gas was observed. We distinguish between targets where only cold CO gas is observed, and targets where both warm and cold temperatures is found. In the latter cases, cold-gas measurements are likely more uncertain due to the contribution of the high-temperature on the low-temperature lines.

\subsection{Solid $\rm{[^{12}C^{16}O}]$/$\rm[{^{13}C^{16}O}]$ toward IRS 51}
The optical depth for the $^{13}$CO ice toward the foreground cloud traced by IRS 51 provides the only solid [$^{12}$C$^{16}$O]/[$^{13}$C$^{16}$O] ratio ($68\pm10$) that can be compared to a gas-phase ratio along the same line-of-sight in this study. The ice-phase ratio is significantly lower than our derived gas-phase ratio of $165\pm10$. The [$^{12}$C$^{16}$O]/[$^{13}$C$^{16}$O] value in ice for IRS 51 is roughly consistent with other published [$^{12}$C$^{16}$O]/[$^{13}$C$^{16}$O] values for ice in YSO targets: \cite{boogert2002a} derive a [$^{12}$C$^{16}$O]/[$^{13}$C$^{16}$O] ratio of $71\pm15$ for the ice observed toward the massive YSO NGC 7538 IRS 9, using Keck-NIRSPEC, and a lower ratio of $45\pm15$ was observed toward the CRBR 2422.8-3423 edge-on disk using VLT-ISAAC \citep{pontoppidan2005}. 

\begin{deluxetable*}{llllll}[ht]
\centering
\tablecolumns{6}
\tablewidth{0pt}
\tablecaption{CO ice profile optical depths, column densities, and ice fractions for the CRIRES targets\tablenotemark{a,b}}

\tablehead{
\colhead{Target} & \colhead{Component} & \colhead{$\tau_{\rm mc}$} & \colhead{[$^{12}$C$^{16}$O]$\rm_{\rm ice}$}   & \colhead{[$^{12}$C$^{16}$O]$\rm_{\rm gas}$}  & \colhead{CO ice fraction}\tablenotemark{c}\\
   & & & \colhead{($10^{18}$ cm$^{-2}$)}  &\colhead{($10^{18}$ cm$^{-2}$)}     &
     }
\startdata 
DoAr24E S\tablenotemark{d} & single & $<0.02$          &$<0.01$           & $1.10\pm 0.005$  & $<0.01$  \\
DoAr24E N\tablenotemark{d} & single & $<0.02$          &$<0.01$           & $1.09\pm 0.003$  & $<0.01$  \\
VV CrA N\tablenotemark{d}  & low    & $<0.05$          &$<0.03$           & $1.41\pm 0.02$   & $<0.02$  \\ 
VV CrA N\tablenotemark{d}  & high   & $<0.05$          &$<0.03$           & $7.62\pm 0.07$   & $<0.004$ \\ 
VV CrA S\tablenotemark{d}  & single & $<0.05$          &$<0.03$           & $1.18\pm 0.004$  & $<0.03$  \\
HL Tau\tablenotemark{d}    & low    & $<0.08$          &$<0.05$           & $1.07\pm 0.02$   & $<0.05 $ \\
HL Tau\tablenotemark{d}    & high   & $<0.08$          &$<0.05$           & $4.47\pm 0.06$   & $<0.01 $ \\
LLN 19\tablenotemark{e}    & single & $0.03\pm 0.002$  & $0.02\pm 0.001$  & $2.02\pm 0.03$   & $0.01\pm 0.001$ \\
RNO 91                     & low    & $0.25\pm 0.01$   & $0.17\pm 0.004$  & $0.69\pm 0.07$   & $0.19\pm 0.005$ \\
RNO 91                     & high   & $0.25\pm 0.01$   & $0.17\pm 0.004$  & $2.36\pm 0.03$   & $0.07\pm 0.002$ \\
Reipurth 50                & low    & $0.51\pm 0.003$  & $0.34\pm 0.002$  & $1.98\pm 0.16$   & $0.15\pm 0.01$  \\
Reipurth 50                & high   & $0.51\pm 0.003$  & $0.34\pm 0.002$  & $10.8\pm 0.77$   & $0.03\pm 0.002$ \\
IRS 63                     & low    & $1.69\pm 0.01$   & $1.12\pm 0.01$   & $1.90\pm 0.03$   & $0.37\pm 0.004$ \\
IRS 63                     & high   & $1.69\pm 0.01$   & $1.12\pm 0.01$   & $5.74\pm 0.15$   & $0.16\pm 0.004$ \\
IRS 51                     & single & $4.54\pm 0.05$   & $3.01\pm 0.03$   & $2.03\pm 0.04$   & $0.60\pm 0.01$  

\enddata
\tablecomments{
\tablenotetext{a}{Optical depths for all targets excepting LLN 19 are derived from the middle component as determined by the phenomenological decompositions in \cite{Pontoppidan2003}.} 
\tablenotetext{b}{Uncertainties are $1\sigma$.}
\tablenotetext{c}{[$^{12}$C$^{16}$O]$\rm_{Ice}$/([$^{12}$C$^{16}$O]$\rm_{Ice}$ + [$^{12}$C$^{16}$O]$\rm_{Gas}$.}
\tablenotetext{d}{Upper limits on the CO ice column from spectra in this paper.}
\tablenotetext{e}{Optical depth is from the middle component values from the phenomenological decompositions in \cite{thi2010}.} 
}
\label{ice_table}
\end{deluxetable*}

\section{Discussion}
\subsection{Galactic $\rm{[^{12}C^{16}O}]$/$\rm[{^{13}C^{16}O}]$ heterogeneity} 
We find significant heterogeneity in gas-phase [$^{12}$C$^{16}$O]/[$^{13}$C$^{16}$O] toward our targets (Figures \ref{gc_co_plot}  and \ref{ice_vs_ratio}), with most values falling above local ISM values of $\sim65-69$, as determined by sub-millimeter observations \citep{Wilson1999, Milam2005}. The large spread in [$^{12}$C$^{16}$O]/[$^{13}$C$^{16}$O] suggests either heterogeneity in the carbon isotopic composition in our local solar neighborhood (e.g. R$_\textrm{\tiny GC}$$\sim8$ kpc), and/or one or more new fractionation mechanisms for CO operating in the dense envelopes of young stars. The distribution of CO isotopologue ratios derived from our CRIRES data do not match the general Galactic trend, which suggests that the carbon reservoirs of young stellar envelopes do not necessarily isotopically represent their parent clouds, as is often assumed in GCE models \citep{Prantzos1996}. 

We are not the first to observe unusual [$^{12}$C$^{16}$O]/[$^{13}$C$^{16}$O] abundance ratios in YSOs and molecular clouds (Figure \ref{gc_co_plot}). In ultraviolet absorption, \cite{lambert1994} and \cite{federman2003} found high [$^{12}$C$^{16}$O]/[$^{13}$C$^{16}$O] values toward diffuse regions in the Ophiuchus cloud, with values ranging from $\sim117$ to 167. These results have been attributed to selective CO photodissociation by external ultraviolet radiation \citep{lambert1994, federman2003}. We discuss connections with CO photodissociation in Section \ref{photodissociation}. 

Unusually high [$^{12}$C$^{16}$O]/[$^{13}$C$^{16}$O] values were also detected in rovibrational absorption using Subaru ($\lambda/\Delta\lambda \approx 23,000)$ toward the massive young stars LkH$\alpha$101 and MonR2 IRS3, with values of $\sim137$ and 158, respectively \citep{goto2003}. \cite{goto2003} suggested that photon-trapping in the rotational lines of the most abundant isotope as a possible explanation for the apparently elevated carbon isotope ratios, although the potential effect was not quantified in their study. 

\subsection{Ice-gas partitioning of $\rm{^{12}C^{16}O}$ and $\rm{^{13}C^{16}O}$}
\subsubsection{Observed relation between the ice-gas fraction and the CO isotopologue ratios}
A possible mechanism for the observed CO fractionation is mass-dependent exchange between gas and ice. Compared to single-dish, beam-averaged millimeter measurements of CO gas-phase isotopologue ratios in molecular clouds, rovibrational absorption through protostellar envelopes is likely to probe denser gas. Dense, cold gas is known to lead to efficient CO freeze-out onto dust grains \citep{jorgensen04}. If the adsorption/desorption rates of CO are mass-dependent, isotopologue fractionation may occur. Further, if the heavier isotopologue ($^{13}$CO) is more tightly bound to the dust grains, we would expect a increase in the [$^{12}$CO]/[$^{13}$CO] ratio in the gas, which is consistent with the detection of large columns of CO ice along the observed lines-of-sight. Here we discuss whether ice-gas partitioning is a possible explanation for the observed high CO isotopologue ratios in the gas. 

The relative fraction of CO sequestered in ice, averaged along the line-of-sight, is plotted against the [$^{12}$C$^{16}$O]/[$^{13}$C$^{16}$O] abundance ratios in the gas (Figure \ref{ice_vs_ratio}). While we might expect to see increased values of the CO abundance ratio for lines-of-sight with a high ice/gas fraction if $^{13}$CO is preferentially bound in the ice mantles, we instead find significant scatter, much larger than the individual errors on each data point, with no simple apparent relation between these quantities. A correlation is particularly absent between cold-gas reservoirs and CO ice fraction, although these two quantities are likely cospatial and trace the same total CO reservoir. However, the line-of-sight with the largest ice fraction (IRS 51) also has the highest CO isotopologue ratio in the gas. Therefore, while other factors are likely equally important, we cannot rule out that ice-gas partitioning could play a significant role in the observed high gas-phase [$^{12}$C$^{16}$O]/[$^{13}$C$^{16}$O].

With respect to other factors affecting the gas-phase ratios, we note that high [$^{12}$C$^{16}$O]/[$^{13}$C$^{16}$O] ratios are also observed in the warm gas in some lines-of-sight (Figure \ref{ice_vs_ratio}, red diamonds), which is unexpected if ice-gas partitioning is the only fractionation mechanism, since environments warm enough to fully sublimate the CO ice should reset the gas isotopologue fraction back to the canonical value of $\sim 65$. In order to reconcile this with ice-gas partitioning, there must be a mechanism to effectively remove CO from the ice during the cold phase of envelope evolution. Chemical pathways removing CO from the ice include the successive hydrogenation of CO to formaldehyde (H$_{2}$CO) and methanol (CH$_{3}$OH) \citep{hidaka2004, watanabe2002, cuppen2009, fuchs2009}, or reactions with OH or O to form CO$_2$ \citep{mennella2006, Oba2010, Ioppolo2011}. Further, we would expect the column densities of ice to be comparable or larger than those of the gas in order for the ice to be a significant reservoir that could increase the gas-phase [$^{12}$C$^{16}$O]/[$^{13}$C$^{16}$O]. For IRS 51, the large discrepancy in [$^{12}$C$^{16}$O]/[$^{13}$C$^{16}$O] between the ice ($\sim68$) and the gas ($\sim{165}$), along with a seemingly low CO ice fraction ($\sim0.60$), suggests instead that open-system chemistry may be occurring via CO ice reactions.  

While the data are inconclusive with regard to carbon isotopic fractionation between reservoirs of CO ice and gas, potential trends in cold- and warm-gas reservoirs can be further investigated with high-resolution spectroscopy for a larger sample of protostellar lines-of-sight. Further, models considering chemical evolution in protoplanetary environments including YSOs could consider interactions between ice and gas reservoirs as a potential influence on isotope partitioning. 

\subsubsection{Comparison to theoretical and experimental studies}
To estimate whether there is an expectation of ice-gas partitioning for CO, we briefly review the theoretical basis for a potential mass-dependence in the ice-gas balance. Following a standard approach for ice-gas partitioning \citep[e.g.][]{tielens1982}, CO separations between the solid- and gas-phase reservoirs can be investigated with respect to mass-dependent kinetics and the balance between the rates of $^{12}$C$^{16}$O and $^{13}$C$^{16}$O gas adsorption and desorption, as given by the relation,

\begin{equation}
\frac{dn_{\rm ice}}{dt}=R_{\rm ads} - R_{\rm des},
\label{eq:rate}
\end{equation}

where $n_{\rm ice}$ is the number density of ice, $R_{\rm des}= \nu_{0}$exp$(-\Delta H/kT)$$\times$ $n$$_{\rm CO,ice}\times\beta$, and $R_{\rm ads}=n_{\rm CO,gas}\times n_{\rm dust}\times\pi d^{2}\times\sqrt{3kT/m_{\rm CO}}\times f$ are the desorption and adsorption rates, respectively. The factor, $\Delta H$, is the binding energy of CO. The factor $\beta$ accounts for CO desorbing only from the top monolayer of the ice, $d$ is the grain radius, and $f$ is the sticking coefficient. The mass of CO $(m_{\rm CO})$ is an obvious mass-dependency of this relation, and $\nu_{0}$ is the pre-exponential factor. The number densities of the CO gas and dust particles are given by $n_{\rm CO, gas}$ and $n_{\rm dust}$, respectively \citep{pontoppidan2006}. In addition to the mass itself, the binding energy and pre-exponential factors are likely both mass-dependent, although to unknown degrees. 

\begin{figure}
\centering
\includegraphics[width=8.5cm]{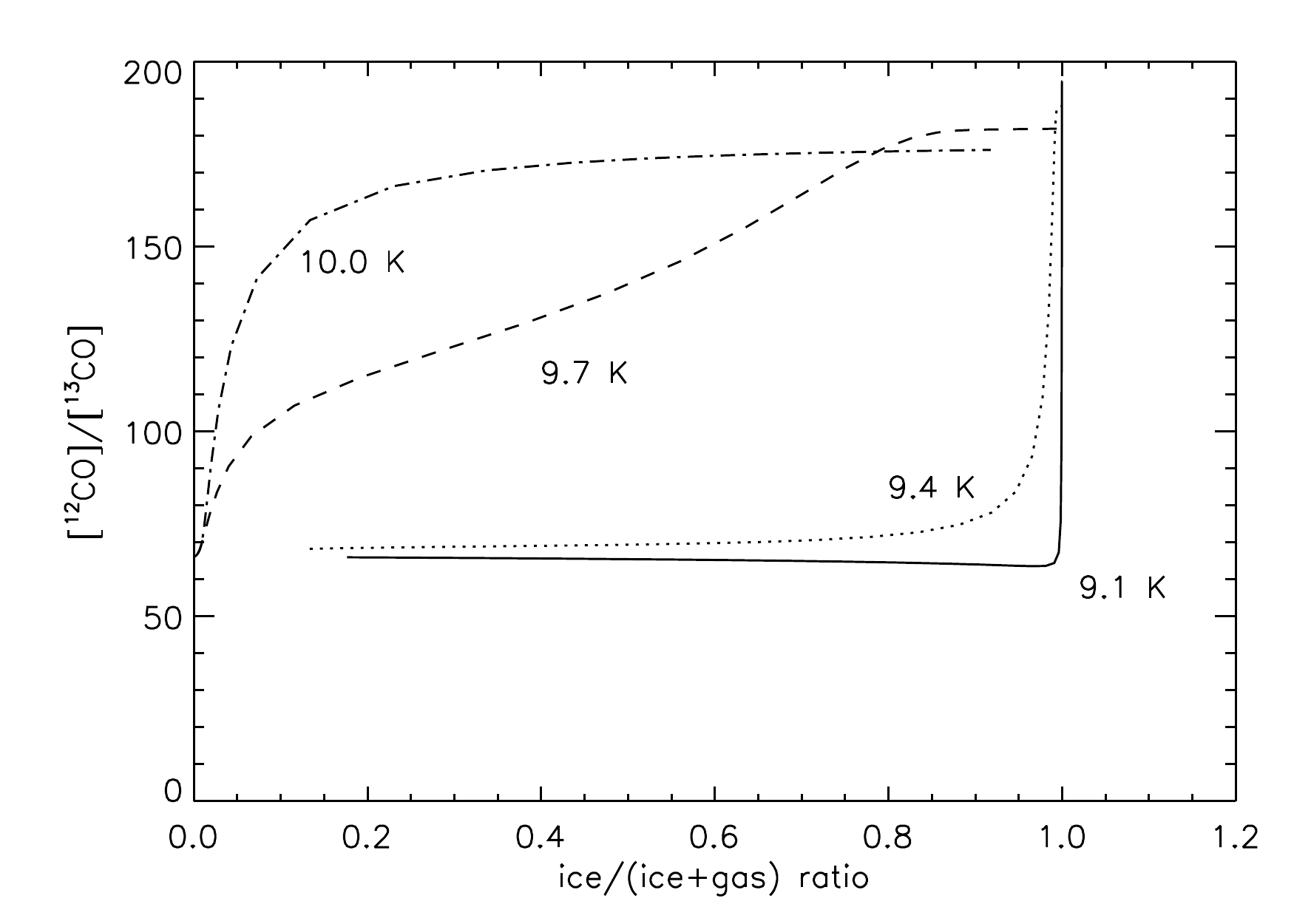}
\caption[]{The gas-phase [$^{12}$CO]/[$^{13}$CO] ratio predicted by the simple model described by Eq. \ref{eq:rate}, assuming a difference in binding energy of 10\,K between the two isotopologues, as allowed by the experimental uncertainty, and a pre-exponential that scales with $m_{\rm CO}^{-1/2}$. The model is a snapshot at $5\times 10^5\,$yr and the individual curves are created by varying the gas density from $10^3$ to $10^6\,\rm cm^{-3}$ and temperature from 9.1 to 10.0\,K, as indicated.}
\label{ratio_model}
\end{figure}

Some experimental work exists for the mass dependence of thermal desorption of CO. In an ultra-high vacuum, temperature-programmed desorption (TPD) experiment, \cite{acharyya2007} measured binding energies and pre-exponentials for desorption from pure $^{12}$CO and $^{13}$CO ices. The parameters were derived using an empirical kinetic model. However, no difference between the $^{12}$CO and $^{13}$CO binding energies could be measured within the experimental uncertainties: $858\pm15$ K and $854\pm10$ K for $^{12}$C$^{16}$O and $^{13}$C$^{16}$O, respectively. 

To estimate the degree of ice-gas fractionation allowed by the boundaries imposed by the experimental evidence, we solved for the ice abundance in Eq. \ref{eq:rate}, assuming a 10\,K difference in binding energy between the two CO isotopologues and a pre-exponential that scales as $m_{\rm CO}^{-1/2}$ \citep{hasegawa1992}. In Figure \ref{ratio_model} it is seen that, under these assumptions, the [$^{12}$CO]/[$^{13}$C] gas ratio may reach values as high as 170, but only for a narrow range of temperatures close to those of CO desorption (above which no ice exists). Essentially, a dust temperature exists at which $^{13}$CO can efficiently remain bound to the dust surface, while $^{12}$CO cannot. Nearly all of this effect is due to the assumed difference of only 10\,K in the binding energies. Compared to what is currently available, significantly more precise laboratory measurements of mass-dependent binding energies would be needed to rule out binding-energy differences as  a significant cause of ice-gas fractionation. While such temperature-dependent fractionation may be consistent with the observed spread in isotopologue ratios, it remains unclear to which degree the physical conditions of real molecular clouds and envelopes must be fine-tuned to lead to significant ice-gas fractionation. 

More realistic modeling of ice-gas fractionation of CO is potentially complicated by several factors. For example, investigations of the effects of CO entrapment in the ice on the desorption and adsorption (both potential mass-dependent processes) in astrophysical environments have shown that diffusion of CO into the porous H$_{2}$O ice leads to competing CO entrapment and diffusion effects on the desorption of CO, the degree to which depends on the ice temperature \citep{collings2003}. Further, a recent study has shown that three-phase modeling of H$_{2}$O:CO$_{2}$ ice desorption depends on the ice thickness and mixing ratios, which may have significant effects on ice desorption \citep{fayolle2011}.

\subsection{Oxygen isotopologues and $\rm{CO}$ self-shielding}
\label{photodissociation}
Isotope-selective photodissociation of CO due to self-shielding refers to the differential shielding of CO isotopologues from photodissociation by far ultraviolet (FUV) radiation in proportion to their column densities. With regard to oxygen isotopes, this leads to spatial regions characterized by more rapid destruction of $^{12}$C$^{17}$O and $^{12}$C$^{18}$O compared to the much more abundant $^{12}$C$^{16}$O molecule. Oxygen liberated during this process is thought to eventually end up in H$_{2}$O, providing an explanation for the overabundance of H$_{2}^{18}$O and H$_{2}^{17}$O relative to H$_{2}^{16}$O in the early solar system. Isotope-selective photodissociation of CO due to self-shielding has been shown to be a relevant process in the interstellar medium \citep{vandishoeck1988, visser2009}. 

The solar system oxygen isotope anomaly is characterized by the mass-independent correlation between [$^{16}$O]/[$^{18}$O] and [$^{16}$O]/[$^{17}$O] found in meteoritic materials as compared to the mass-dependent distribution known in terrestrial rocks. Mass independence refers to relative differences in [$^{16}$O]/[$^{18}$O] and [$^{16}$O]/[$^{17}$O] among primitive rocky objects that are nearly identical, suggesting changes in [$^{16}$O] relative to both [$^{17}$O] and [$^{18}$O], rather than the expected mass-dependent trend in which relative changes in [$^{16}$O]/[$^{17}$O] are about half those in [$^{16}$O]/[$^{18}$O] on a log scale. The solar system oxygen anomaly has defied conclusive explanation since its discovery \citep{Clayton1973}, and has been investigated observationally by \cite{brittain2005} and \cite{smith2009}, the latter study finding mass-independent signatures in oxygen isotopes observed in CO toward the VV CrA N disk, consistent with isotope-specific photodissociation by CO self-shielding. 

\begin{figure}[ht]
\centering 
\includegraphics[width=8.5cm]{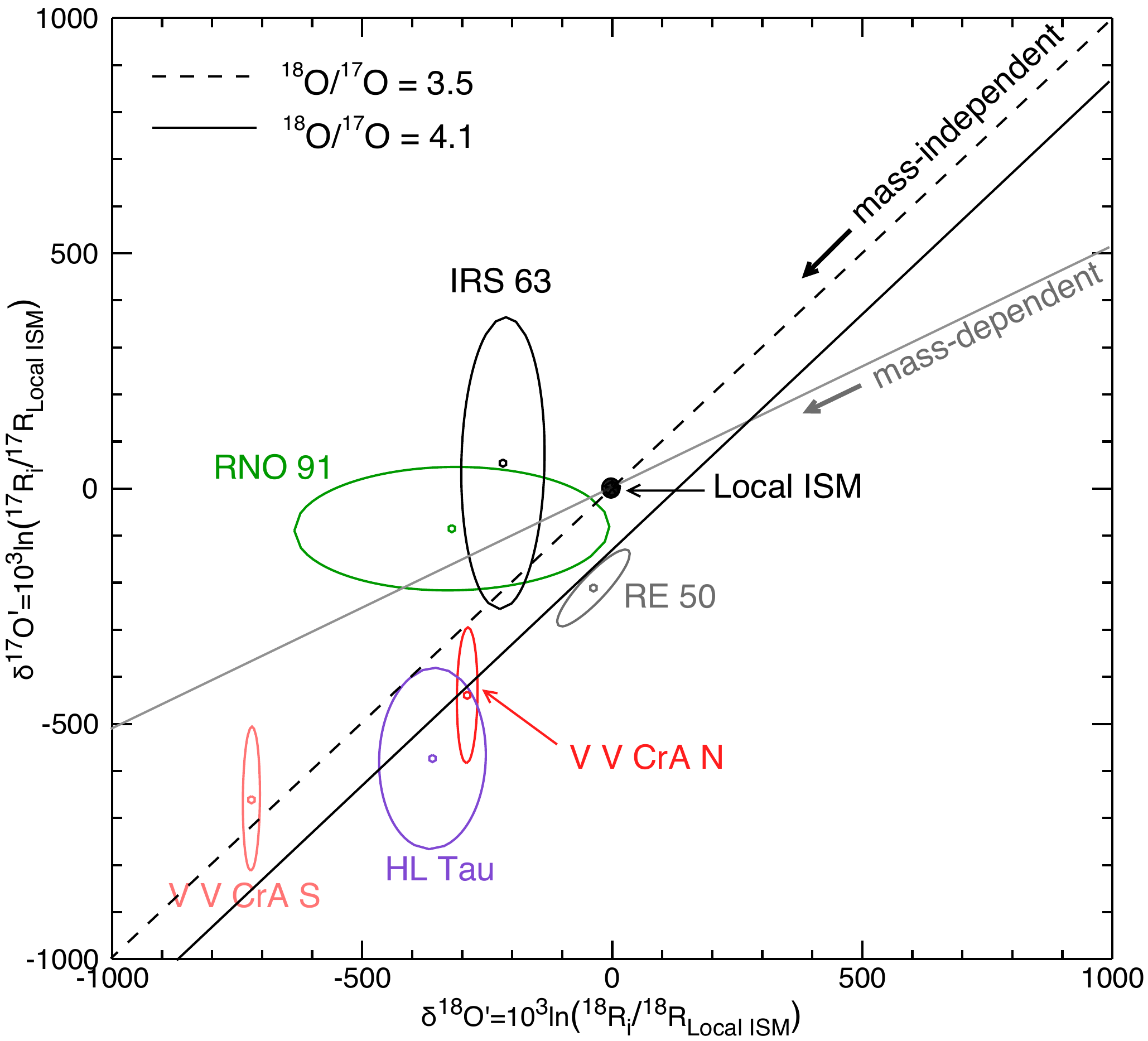}
\caption[]{Comparison of oxygen isotope ratios between the interstellar medium and CO surrounding the targets in our study where 3 oxygen isotopes were observed. Oxygen isotope ratios are plotted in linearized delta notation, such that the [$^{16}$O]/[$^{17}$O] and [$^{16}$O]/[$^{18}$O] from local ISM values are shown in per mil using the linearized form of the delta notation. Here, $\delta^{18}{\rm O}^{\prime} = 10^3
\ln(^{18}R_i/^{18}R_{\rm Local\ ISM})$ 
and $^{18}R_i$ = [C$^{18}$O]/[C$^{16}$O]. Ratios for the YSOs are derived from the best-constrained (high- or single-T) values from the rotational analyses (Table \ref{ratio_table}). The solid and dashed lines show mass-independent fractionation (slope $\sim1$)
lines with different assumptions for the $^{18}$O/$^{17}$O ratio, as indicated. The mass-dependent line (slope $\sim1/2$) is also shown for comparison. Ellipses represent 65\% confidence limits (standard deviation) derived from $1\sigma$ errors from the $\tau_{\circ}$ for each line. Targets are colored according to parent cloud (see also Figure \ref{gc_co_plot}).}
\label{delta_delta}
\end{figure}

The compiled oxygen data for this study, which also includes refined analyses of VV CrA N and Reipurth 50 from \cite{smith2009}, are shown in Table \ref{ratio_table} and Figure \ref{delta_delta}. The mass-dependent lines in Figure \ref{delta_delta} are derived by estimating the oxygen isotope ratios for the ISM as a function of distance from the Galactic center. These values can be estimated from the 4 kpc and 8 kpc values \citep{Wilson1999}, such that for $\rm [^{16}O]/[^{18}O]$:

\begin{equation}
{\rm[ ^{16}O]/[^{18}O]}_{R_{\rm GC}} = (57.5 \pm 10)\  {R_{\rm GC}} + 97,
\label{eq:5}
\end{equation}

\noindent where the $1\sigma$ uncertainty in the slope reflects the uncertainty in the standard errors reported in Table 4 of \cite{Wilson1999}. For $\rm [^{16}O]/[^{17}O]$, equation \ref{eq:5} 
is combined with values for $\rm [^{18}O]/[^{17}O]$ in the ISM, which vary from 3.5 \citep{Wilson1999} to 4.1 \citep{Wouterloot2008} (Figure \ref{delta_delta}).

We find a mass-independent signature in the oxygen isotopes relative to solar system abundances, consistent with CO self-shielding, toward the HL Tau disk, and the same mass-independent signature is retrieved in the refined results for VV CrA N (as compared to \cite{smith2009}). These findings lend further support for models considering CO self-shielding on disk surfaces \citep{lyons2005, young2008}. It is particularly interesting to note that the recent suggestions of planet formation in the HL Tau disk based on new ALMA imaging \citep{tamayo2015} would imply that CO self-shielding may be a process that occurs and/or persists later in the evolution of YSOs than previously thought.

We would not expect UV photolysis to be a significant process in the foreground cloud traced by VV CrA S, or in the relatively embedded objects, RNO 91, Reipurth 50, and IRS 63. Nevertheless, due to the errors in the rare isotopologue, $^{12}$C$^{17}$O, a mass-dependent effect cannot be distinguished from mass-independence in oxygen fractionation for RNO 91 and IRS 63. Additional high-quality CO data for YSOs would help in evaluating mass-independence in oxygen for objects in various stages of evolution. 

With regard to the rare CO isotopolgues, with the exception of low-temperature gas in Reipurth 50, the [$^{12}$C$^{18}$O]/[$^{12}$C$^{17}$O] ratios are within $1-2\sigma$ of molecular cloud radio measurements for local ISM, further strengthening the claim that the solar system is anomalous in $^{18}$O/$^{17}$O with respect to the local solar neighborhood \citep{young2011}. This consistency in our data is observational support for $^{12}$C$^{18}$O and $^{12}$C$^{17}$O being similarly affected in young stellar environments. 

\subsubsection{Evaluation of $\rm{CO}$ self-shielding on carbon isotope ratios}
\begin{figure}[ht]
\centering
\includegraphics[width=8.5cm]{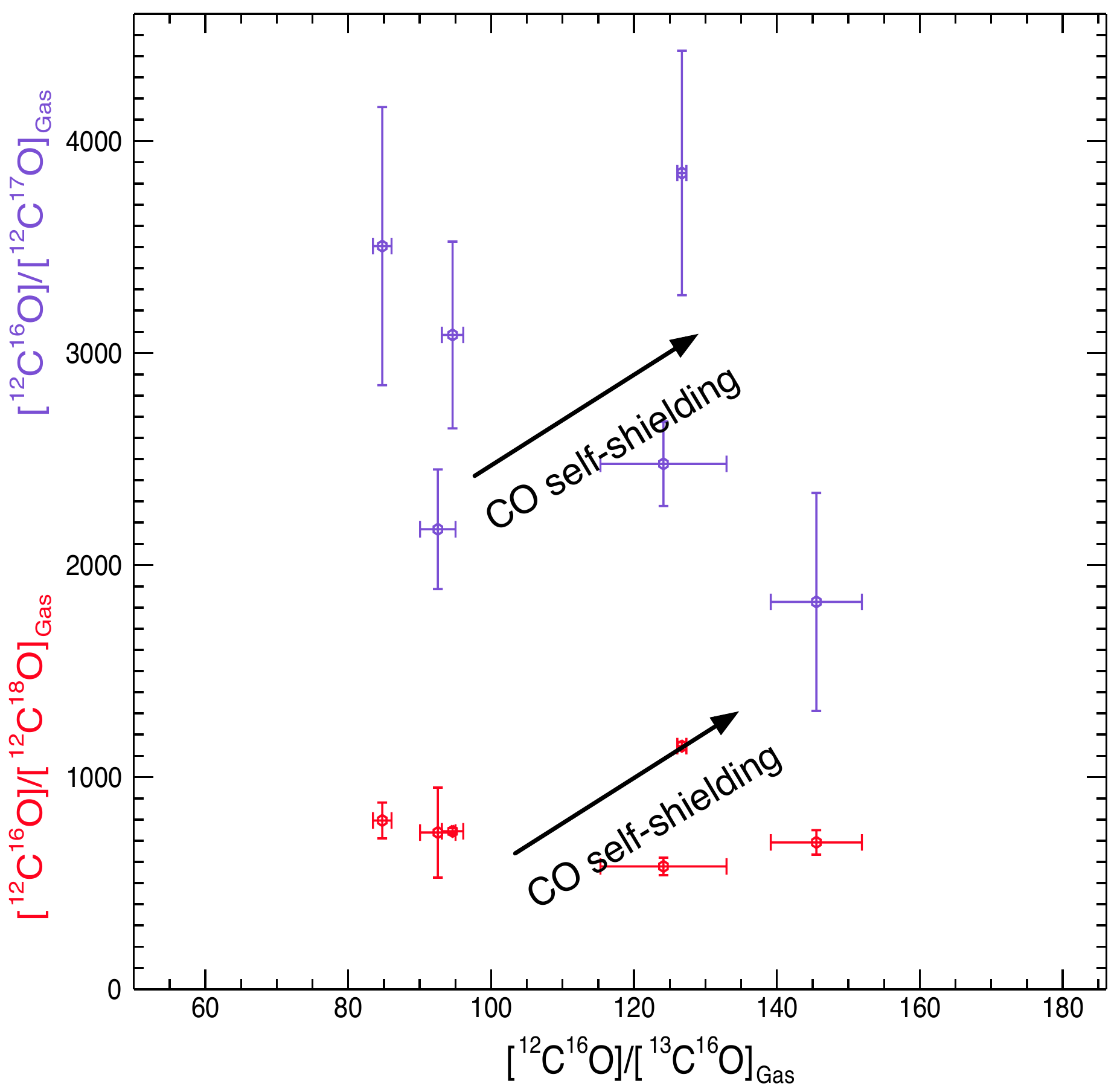}
\caption[]{Carbon isotope ratios against oxygen isotopic abundances in targets where all isotopologues of CO were observed. The [$^{12}$C$^{16}$O]/[$^{12}$C$^{17}$O] ratio is shown in purple, and  [$^{12}$C$^{16}$O]/[$^{12}$C$^{18}$O] in red.  The derived [$^{16}$O]/[$^{17}$O] changes by nearly a factor of two relative to [$^{12}$C$^{16}$O]/[$^{13}$C$^{16}$O], while the [$^{16}$O]/[$^{18}$O] varies by roughly 1.5 times with [$^{12}$C$^{16}$O]/[$^{13}$C$^{16}$O]. The arrows indicate the relative trends we would expect if isotopologue-specific photodissociation by CO self-shielding were the predominant process affecting the oxygen and carbon isotope ratios.}
\label{oxygen_carbon}
\end{figure}

While critical evaluation of mass-independence requires three isotopes, we qualitatively considered if isotope-selective photodissociation by CO self-shielding could be affecting the derived [$^{12}$C$^{16}$O]/[$^{13}$C$^{16}$O] toward our targets by comparing the oxygen and carbon isotope ratios where this was possible: VV CrA N, VV CrA S, HL Tau, RNO 91, Reipurth 50, and IRS 63. These are plotted in Figure \ref{oxygen_carbon}, with the qualitative relative trend for CO self-shielding overplotted on the data. The derived [$^{16}$O]/[$^{17}$O] changes by nearly a factor of two relative to [$^{12}$C$^{16}$O]/[$^{13}$C$^{16}$O], while the [$^{12}$C$^{16}$O]/[$^{12}$C$^{18}$O] varies by roughly 1.5 times with [$^{12}$C]/[$^{13}$C]. If CO self-shielding were the dominant mechanism, we would expect at least positive, direct changes in the carbon with the oxygen isotopes along an integrated line-of-sight, as well as similar variability in [$^{12}$C$^{18}$O] and [$^{12}$C$^{17}$O] compared with [$^{12}$C$^{16}$O]/[$^{13}$C$^{16}$O] due to their similar rates of photodissociation \citep[e.g.][]{vandishoeck1988, visser2009}. Therefore, while we cannot be certain that photodissociation of CO has no affect on the [$^{12}$C$^{16}$O]/[$^{13}$C$^{16}$O], we cannot claim that it is a dominant process affecting this carbon reservoir. 

Models for low-mass YSOs predict that [$^{12}$C$^{16}$O]/[$^{13}$C$^{16}$O] should not increase significantly due to photodissociation \citep[e.g.][]{visser2009, Woods2009}. Atomic carbon liberated by CO photodissociation is photoionized largely to C$^+$, and C$^+$ reacts rapidly to exchange carbon with CO by the reaction $^{13}$C$^+$ + $^{12}{\textrm{C}}$O $\rightleftarrows$ $^{12}$C$^+$ + $^{13}$CO + 35 K \citep{Warin1996, vandishoeck1988, Langer84}, driving $^{13}$C back into CO and diminishing the signature of selective photodissociation. 
However, very high [$^{12}$C$^{16}$O]/[$^{13}$C$^{16}$O] ratios derived for diffuse clouds have nevertheless been attributed to CO self-shielding. Values of $\sim117$, $125$ \citep{federman2003}, and $164$ \citep{lambert1994} have been observed toward diffuse Ophiuchus clouds that are thought to be subjected to high-UV fields, and where CO photodissociation is thought to dominate over the carbon exchange reaction that would otherwise erase a photochemical signature. 

While the upper range in [$^{12}$C$^{16}$O]/[$^{13}$C$^{16}$O] in our sample is consistent with the Ophiuchus data (Figure \ref{gc_co_plot}), we do not find the expected mass-independent signatures in oxygen isotopes consistent with CO self-shielding. Further, as compared to diffuse clouds, we would not expect the molecular gas deep inside embedded YSO envelopes to be subjected to such high UV fields given their relatively high values of visual extinction ($A_{v}\sim10$ \citep[e.g.][]{vuong2003}). 

\section{Conclusions}
Heterogenous [$^{12}$C$^{16}$O]/[$^{13}$C$^{16}$O] gas-phase abundance ratios, with several unusually high values, have been found using infrared absorption spectroscopy at very high resolution toward a sample of YSOs in nearby star-forming clouds. A possible partial explanation for this trend may be CO isotopologue partitioning between the ice and gas reservoirs. The degree of partitioning and the preservation of the relation in warm regions where the CO ice has sublimated would require the action of a process that can remove CO from the ice phase without returning it to the gas. Additional high-quality data could help substantiate these findings. 

Signatures consistent with isotopologue-selective photodissociation by CO self-shielding in oxygen were observed in the oxygen isotopes toward the disks, HL Tau and VV CrA. However, our results do not support CO self-shielding as a dominant mechanism for explaining the trends in the gas-phase [$^{12}$C$^{16}$O]/[$^{13}$C$^{16}$O] heterogeneity. 

We conclude that chemical evolution models for young stellar environments should consider the interaction between CO ice and gas reservoirs as a potential fractionation mechanism affecting the carbon isotope budget and influencing carbon isotopic heterogeneity in the gas. Interactions between these reservoirs could also contribute to isotopic heterogeneity between solar-type YSOs and their parent clouds. Finally, isotopologue-selective photodissociation by CO self-shielding should be considered an important mechanism affecting oxygen isotopes in protoplanetary disks, potentially persisting in the later stages of planet formation.

\acknowledgments{This research was supported in part by a grant from the NASA Origins Program (EDY, MRM, RLS). This work is based on observations collected at the European Southern Observatory Very Large Telescope under program ID 179.C-0151. Support from the STScI Directors Discretionary Fund is gratefully acknowledged. }

\bibliographystyle{apj}

\end{document}